%% file: main.tex
\documentclass{article}
\PassOptionsToPackage{numbers, compress}{natbib}
\usepackage[preprint]{neurips_2025}
\usepackage[utf8]{inputenc}
\usepackage[T1]{fontenc}
\usepackage{amsmath, amssymb}
\usepackage{graphicx}
\usepackage{caption}
\usepackage{booktabs}
\usepackage{microtype}
\usepackage{siunitx}
\usepackage{pdfpages}
\usepackage{xcolor}
\usepackage{placeins}
\usepackage[colorlinks]{hyperref}

\title{Alljoined-1.6M: A Million-Trial EEG-Image Dataset for Evaluating Affordable Brain-Computer Interfaces}
\author{%
  \textbf{Jonathan Xu}$^{1}$\quad
  Ugo Bruzadin Nunes$^{1}$\quad
  Wangshu Jiang$^{1,2}$\quad
  \textbf{Samuel Ryther}$^{1}$\quad
  \textbf{Jordan Pringle}$^{1}$\\
  \textbf{Paul S.\ Scotti}$^{3,4}$\quad
  \textbf{Arnaud Delorme}$^{5}$\quad 
  \textbf{Reese Kneeland}$^{1}$\quad
  \\
  $^{1}$Alljoined\\
  $^{2}$University of Waterloo\\
  $^{3}$Sophont \\
  $^{4}$Princeton Neuroscience Institute \\
  $^{5}$University of California San Diego
}
\begin{document}
\maketitle

\begin{abstract}
We present a new large-scale electroencephalography (EEG) dataset as part of the THINGS initiative, comprising over $1.6$ million visual stimulus trials collected from 20 participants, and totaling more than twice the size of the current most popular benchmark dataset, THINGS-EEG2. Crucially, our data was recorded using a 32-channel consumer-grade wet electrode system costing \(\sim\)\$2.2k - around 27x cheaper than research-grade EEG systems typically used in cognitive neuroscience labs. Our work is one of the first open-source, large-scale EEG resource designed to closely reflect the quality of hardware that is practical to deploy in real-world, downstream applications of brain-computer interfaces (BCIs). We aim to explore the specific question of whether deep neural network-based BCI research and semantic decoding methods can be effectively conducted with such affordable systems—filling an important gap in current literature that is extremely relevant for future research. In our analysis, we not only demonstrate that decoding of high-level semantic information from EEG of seen images is possible at consumer-grade hardware, but also that our data can facilitate effective EEG-to-Image reconstruction even despite significantly lower signal-to-noise ratios. In addition to traditional benchmarks, we also conduct analyses of EEG-to-Image models that demonstrate log-linear decoding performance with increasing data volume on our data, and discuss the trade-offs between hardware cost, signal fidelity, and the scale of data collection efforts in increasing the size and utility of currently available datasets. Our contributions aim to pave the way for large-scale, cost-effective EEG research with widely accessible equipment, and position our dataset as a unique resource for the democratization and development of effective deep neural models of visual cognition.
\end{abstract}

\section{Introduction}
\label{sec:intro}
Electroencephalography (EEG) is a widely used non-invasive method to measure human brain activity, with applications ranging from fundamental neuroscience to advanced brain-computer interfaces (BCI). A longstanding challenge in the development of BCIs is the trade-off between data quality and accessibility: high-density research-grade EEG systems provide high signal fidelity but are prohibitively expensive and resource-intensive, limiting the scope of data collection and the practicality of real-world applications. Consequently, most EEG studies rely on relatively small datasets, and real-world applications are rare. Despite these current limitations, recent advances in affordable EEG hardware and machine learning algorithms to decode brain data have begun to change this equation. 

The cost of consumer-grade EEG systems has declined substantially, potentially lowering the barrier to BCI applications in medical and consumer sectors, as well as unlocking larger-scale neural data collection efforts. Devices such as the Emotiv EPOC and Flex series are available at a fraction of the cost of traditional research-grade systems, making them appealing for these scenarios. However, these affordable systems generally suffer from reduced signal-to-noise ratio (SNR) and other technical limitations \citep{sabio2024scoping, mikhaylov2024comparison}, and so research-grade EEG remains the standard for most current research efforts and datasets. Nevertheless, an important open question remains: \textit{Can consumer-grade EEG systems facilitate shallow and deep neural decoding efforts?} 

In this work, we aim to answer this question by combining the advantages of scale and accessibility. We introduce a new EEG dataset that is, to our knowledge, the largest of its kind, comprising more than twice the number of subjects as the previously largest human EEG object recognition dataset, THINGS-EEG2 \citep{Gifford2022}. Unlike THINGS-EEG2, which used a 64-channel research-grade EEG system, our dataset was collected with the Emotiv Flex 2, a 32-channel wireless headset retailing at roughly \(\sim\)\$2.2k at the time of writing. Despite its lower cost, the Flex 2 can deliver full-scalp coverage and up to 256 Hz sampling (sub-4 ms temporal precision), enabling many experiments that historically required \(\sim\)\$35 – \$60k research-grade setups. Although the Flex 2 yields lower signal fidelity than research-grade systems, we demonstrate that it is nevertheless useful for facilitating downstream decoding tasks, such as semantic meta-category decoding, image retrieval, and EEG-to-Image reconstruction using state-of-the-art ML models. 

Our contributions are three-fold:
\begin{enumerate} 
\item \textbf{Dataset:} We release Alljoined-1.6M as part of the THINGS initiative, a large scale EEG dataset of visual perception containing 32-channel recordings from 20 participants, 4 sessions each, totaling 1.6 M trials across 16,740 unique images, all collected on affordable EEG hardware (Fig. \ref{figure:header}A,B).
\item \textbf{Benchmarks and Decoding Analysis:} We provide extensive benchmarks and analysis of existing EEG-to-Image reconstruction, retrieval, and semantic decoding models on our dataset, setting a baseline for future research developing methods for decoding affordable EEG responses to visual stimuli.
\item \textbf{Scaling and Cost Analysis:} We conduct a detailed analysis of the within-subject scaling properties of our dataset, finding that, despite the lower SNR, decoding performance still increases log-linearly no signs  of saturation, demonstrating that scaling is still an effective approach for improving decoding performance on consumer-grade EEG hardware. We also demonstrate (in Fig. \ref{figure:header}C) the degree of financial investment necessary to obtain certain benchmarks for decoding performance.
\end{enumerate}

Our findings demonstrate the growing potential of more cost-effective EEG headsets that mirror the hardware constraints of many real-world BCI deployments. By publicly releasing Alljoined-1.6M\footnote{\url{https://huggingface.co/datasets/Alljoined/Alljoined-1.6M}}\footnote{\url{https://github.com/Alljoined/Alljoined-1.6M}} and demonstrating its utility, we hope to democratize progress in EEG-based machine learning and lower the entry barrier for research groups with limited resources.

\section{Related Work}\label{sec:related}

\textbf{Low-Cost EEG Hardware.} Consumer-grade EEG headsets from providers like Emotiv and OpenBCI have dramatically lowered the cost and logistical barriers to neural recording, enabling at-home and mobile experiments. Although these devices typically offer fewer channels and lower signal fidelity than clinical systems, a growing body of work demonstrates that even low-density EEG can yield meaningful biomarkers \citep{badcock2015validation, bhavnani2022acceptability, knierim2025advancing}. For example, Duvinage et al. \citep{duvinage2012p300} compared a 14-channel Emotiv EPOC headset against a 128-channel ANT medical-grade system on a P300 speller task. Although the Emotiv under-performed in single-trial classification, they concluded that the Emotiv could reliably support non-critical applications.

\textbf{Deep Learning for Neural Decoding.} Recent advances in deep learning—most notably transformer architectures \citep{vaswani2017attention}, denoising diffusion probabilistic models \citep{ho2020denoising}, and contrastive language–image pretraining (CLIP) \citep{radford2021learning}—together with the availability of large-scale datasets of human brain activity—have catalyzed a new wave of neural decoding studies \citep{ozcelik2023braindiffuser, scotti_reconstructing_2023,Scotti2024MindEye2,takagi2022_decoding,takagi2023improving, kneeland_brain-optimized_2023, kneeland2023reconstructing, kneeland_second_2023, Li2024, ENIGMA, Fei2024-perceptogram, Song2024}. These approaches learn rich spatiotemporal representations from neural recordings, enabling the decoding of high-level semantic information that eluded earlier methods of analysis such as the study of event-related potentials \citep{luck2014introduction}, EEG topographic analysis \citep{michel2018eeg}, and frequency-band metrics \citep{pfurtscheller1999event}. Such semantically informed decoding frameworks hold great promise for downstream brain–computer interface applications.

\textbf{Scaling Laws in Neuroimaging.} Recent research has demonstrated that scaling up neural data collection efforts can log-linearly improve modeling and decoding performance \cite{banville2025scaling,sato2024scaling}, echoing trends observed in computer vision and natural language processing. Many advances in computer vision have been driven by massive, high-variance image corpora such as ImageNet, which contains over a million labeled images spanning a thousand object categories \citep{deng2009imagenet}. Inspired by this success, the neuroimaging community has pursued analogous scaling: MEG repositories like OMEGA and Cam-CAN amass hundreds of hours of data across dozens of participants \citep{niso2016omega,taylor2017cambridge}; fMRI collections such as BOLD5000 and Generic Object Decoding sample tens of thousands of distinct stimuli \citep{chang_bold5000_2019,horikawa2017generic}; and the Natural Scenes Dataset comprises \(\sim\)$30,000$ unique images viewed over 40+ 7T sessions at an estimated cost of \$450k \citep{allen_massive_2022}. Together, these efforts underscore a field-wide push for larger, more diverse datasets to deepen our understanding of brain representations and to power robust, generalizable models across modalities.

\textbf{EEG Datasets and Benchmarks.} Despite growing interest in deep learning for EEG across both decoding and representation learning \citep{Roy2019,Lawhern2018} domains, public EEG repositories remain overwhelmingly tailored to behavioral paradigms (e.g., motor imagery, sleep staging) or clinical biomarkers. Although there are available large‐scale collections such as the TUH EEG Corpus with clinical recordings \citep{obeid2016temple}, and BCI competitions centered on specific paradigms \citep{tangermann2012review}, neither facilitates high‐level semantic decoding. While initial efforts to collect EEG datasets in response to visual stimuli were revealed to have confounds in the block-design paradigm that allowed high-capacity models to exploit low-frequency drifts rather than genuine visual features \citep{Spampinato2017, li2018training, li2020perils}, more recent contributions from the THINGS initiative have made strides in increasing the experimental rigor of such datasets: THINGS-EEG1 captured $22,248$ unique images across 50 subjects \citep{Grootswagers2022}, and THINGS-EEG2 recorded \(\sim\)$82,350$ trials per participant over $16,740$ stimuli with a 64-channel lab-grade system \citep{Gifford2022} and a shuffled block design with no overlap between training and testing image classes. While these datasets have been massively successful in enabling researchers to decode semantic content from EEG brain activity, they still rely on research-grade hardware, and often fall short of the scale and hardware accessibility constraints necessary to develop for robust, consumer-facing brain-computer interfaces. To address this gap, we release a multi-subject, million-trial EEG corpus collected entirely with a \(\sim\)\$2.2k, 32-channel consumer headset, along with code and benchmarks to enable semantic decoding research at scale.

\section{Alljoined-1.6M}
\label{sec:methods}
\vspace{-10pt}
\begin{figure}[ht]
\centering
\includegraphics[width=\textwidth]{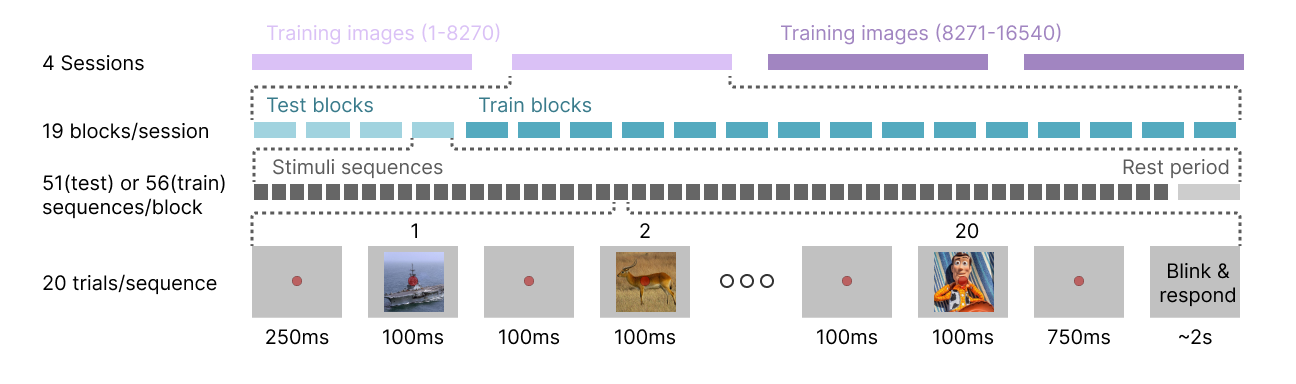}
\vspace{-15pt}
\caption{Experimental paradigm for Alljoined-1.6M. Details can be found in Section \ref{paradigm}.}
\vspace{-5pt}
\label{figure:paradigm}
\end{figure}

% Need to specify reference electrode?
\textbf{Hardware and Recording Setup.} Alljoined-1.6M was collected using the Emotiv Flex 2 EEG system with sintered silver/silver-chloride (Ag/AgCl) gel-based electrodes. The Flex 2 supports 32 EEG channels—we configured a montage covering primarily occipital regions associated with visual perception (detailed in Appendix \ref{app:datadetails}). The Flex 2 streams data wirelessly via Bluetooth 5.2, at a sampling rate of 256 Hz (sufficient to capture event-related potentials). Notably, the entire hardware setup (headset, sensors, software) cost only \(\sim\)\$2.2k, in contrast to the research-grade 64-channel ActiChamp amplifier and cap used in THINGS-EEG2 which we estimated to cost \(\sim\)\$60k. Participants wore the Flex 2 in a dark and quiet room, positioned $60$cm from the screen displaying the stimulus images, with a viewing angle of $7^{\circ}$. Throughout each sequence (including blank trials), a small semi-transparent red fixation dot with a black border ($0.2^{\circ}$ × $0.2^{\circ}$, 50\% opacity) was present at the center of the stimulus. Stimuli were shown against a gray background with an RGB value of (127,127,127), and were presented with PsychoPy. Millisecond-accurate triggers delivered through the Emotiv API aligned the onset of a stimulus image with the timeline of EEG acquisition.
%\ref{app:setup}

\textbf{Participants.}
We recruited 20 healthy adult volunteers (ages 23-63, 15 male, 5 female) from local recruiting platforms in San Francisco, and filtered for participants who had high behavioral scores and high task engagement (from an original pool of 48 subjects). All had normal or corrected-to-normal vision, were provided written informed consent, and were compensated for four recording sessions scheduled on separate days. This approach follows the precedent of NSD \citep{allen_massive_2022} and THINGS-EEG2 \citep{Gifford2022} in obtaining repeated measurements for matched stimuli across multiple recording sessions. To ensure participant safety and to make potentially confounding variables explicit for downstream analyses, we employed two electronic questionnaires whose (anonymized) responses are distributed with the dataset. The screening form gathers stable participant traits—demographics, medical and neurological history, sensory status, prior neuro-imaging, neurodivergence, and data-use consent—to confirm eligibility and document potential confounds. The pre-session questionnaire records transient state variables before each visit (recent caffeine, alcohol or drug use, sleep, fatigue, and meal timing/content) so that session-specific physiological factors can be modeled or controlled. Together, these forms provide a transparent account of both stable (trait) and fluctuating (state) factors for every participant and session, enhancing the reproducibility and secondary-analysis value of the dataset.

\textbf{Stimuli and Experimental Design.}
\label{paradigm}
We used a rapid-serial visual-presentation (RSVP) paradigm paired with an orthogonal target-detection task to keep participants engaged (Fig. \ref{figure:paradigm}). Each trial consists of an image presented for 100ms, followed by a 100ms blank screen. All stimuli were drawn from the THINGS database \citep{Hebart2019}, which contains 26,000 high-resolution photographs spanning 1,854 everyday object categories, and our experiment uses the same set of $16,740$ stimuli utilized in THINGS-EEG2 \cite{Gifford2022} to facilitate direct comparison. Importantly, there is no overlap in the images or image categories presented in the training and test partitions of the dataset, which helps reduce experimental confounds and downstream model overfitting. Each participant completed four recording sessions lasting \(\sim\)2 hours each, with each session comprising 19 RSVP blocks lasting \(\sim\)5 min each. The first 4 blocks of every session presented images from the 200 held-out test images shown in 51 RSVP sequences of 20 images per run, totaling \(4 \times 51 \times 20 = 4,080\) trials for the test data. The remaining 15 blocks of each session presented the remaining $16,540$ training images, randomly split into two equal subsets that were displayed in sessions 1–2 and 3-4. Each test block consisted of 56 RSVP sequences of 20 images. Within a session, every image in the first subset was shown twice, giving four presentations across the two sessions. Images were randomized independently within each session, with the constraint that no image could repeat after fewer than two intervening items (i.e.\ an ABA or AA pattern was disallowed). To encourage vigilance without biasing perception toward any object category, our experiment also included attention check trials. At the end of each sequence, participants given up to 2s to press a key for whether they saw the catch trial of \emph{Woody} appearing in (\(\approx\!6\%\) of images). Performance on these attention trials is described in Appendix \ref{app:behavioral_perf}. A video recording of the stimulus presentation, is also provided for reference on our Huggingface dataset.

\textbf{Dataset Scale.}  
Across the four sessions, each of the $20$ participants completed \(4 \times 20,880 = 83,520\) image trials, resulting in a total dataset size of \(\sim\)$1.6$M trials. Training images were repeated 4–5 times per participant, whereas each test image was shown 80 times, permitting a within-subject averaging procedure to be performed during inference to increase SNR. Total on-task recording time per participant was \(\sim\)8 hours, punctuated by brief breaks.

\textbf{Data Processing and Format.} Raw EEG was stored in standard \texttt{.edf} files and pre-processed with MNE-Python \citep{Gramfort2013}. The Emotiv firmware first applies a dual 50/60 Hz notch filter, and continuous recordings were then epoched from -200 ms to 1000 ms relative to image onset. Synchronization mismatches in the Emotiv trigger stream led us to discard 0–0.6\% of trials between all subjects, which was comparable to exclusion rates reported in earlier Emotiv evaluations \citep{badcock2013validation,williams2020validation}. Epochs were baseline-corrected to the 200 ms pre-stimulus window and resampled to 250 Hz to match the format of the THINGS-EEG2 benchmark \citep{Gifford2022}. As a final preprocessing step, we performed multivariate noise normalization \citep{guggenmos2018multivariate}, to increase SNR, estimating the whitening matrix solely on the training partition to avoid training-test contamination.

\textbf{Meta-Category Groupings.} \label{supercategories} The original THINGS-EEG2 dataset, with its 1,854 fine-grained object categories, has been invaluable for benchmarking deep neural networks but poses challenges for simpler machine learning models and traditional ERP analyses, which often perform better on coarser distinctions facilitating insight into underlying brain activity. These simpler models are critical for leveraging low-cost EEG data in low-resource or real-time settings. Prior work has largely focused on the test set alone, without leveraging the train-test split structure to evaluate generalization across broader semantic boundaries from trained to unseen test images \citep{Gifford2022, Song2024}. To address this, we categorize all trials of our dataset into seven broad meta-category groupings—\textit{Animals},  \textit{Foods/Plants},  \textit{Vehicles},  \textit{Tools}, \textit{Furniture/Household Items},  \textit{Body Parts/Clothing},  and \textit{Toys/Games/Musical Instruments}. For details on how these categories were created, see Appendix \ref{app:statisticalsig}. Our meta-categories are distributed consistently across training and test sets, enabling more interpretable classification and testing of generalization while preserving the dataset’s fine-grained image metadata.
%\ref{app:supercategories}

\section{Analysis and Preliminary Results}
\label{sec:analysis}

We conducted a series of analyses to characterize the dataset and to evaluate the central question of our paper: can current consumer grade hardware facilitate meaningful downstream neural decoding research? We conduct a series of analyses across a wide range of downstream tasks, including semantic decoding, image retrieval, and EEG-to-Image reconstruction tasks.

\begin{figure}[!htb]
    \vspace{-5pt}
    \centering
    \includegraphics[width=\linewidth]{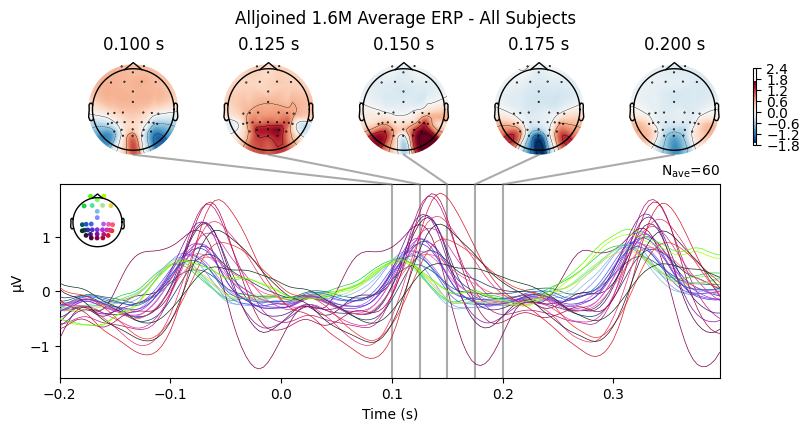}
    \vspace{-10pt}
    \caption{Average Event Related Potential (ERP) across all 20 subjects and all 4 sessions for a total of 1.6 million trials. Topographical maps show changes in visual cortex activity as expected for an RSVP experiment, with primary activity occurring in the occipital cortex for the duration of the image presentation. Three peaks show the ERP peaks of the 200ms interval between image display, essentially one peak per image shown.}
    \vspace{-5pt}
    \label{fig:ERP}
\end{figure}

\textbf{ERP Analysis.} To first demonstrate the reliability of the signal in this dataset, we visually inspected all subjects' session-wise and block-wise ERPs. We observed a pattern similar to the pattern shown in Fig. \ref{fig:ERP} which is typical of a 200ms interval RSVP experiment, with a \(\sim\)100ms peak (P1) and a \(\sim\)200ms trough N200 \citep{luck2014introduction}. Then, we ran a non-parametric spatio-temporal cluster-based permutation test \citep{maris2007nonparametric} to identify clusters of time points and electrodes where condition-specific ERPs diverged significantly across semantic categories. This was performed on averaged ERPs for each condition across trials, and differences were evaluated across subjects. Surprisingly, we found that 16 out of all 21 possible category comparisons yielded significant clusters (\textit{p} < 0.01). The fact that we observe robust category-selective effects under these conditions—using low-cost, consumer-grade EEG hardware—underscores both the sensitivity of the paradigm and the potential suitability for such data for scalable, real-world BCI research. Cluster results are described in more detail in the Appendix \ref{app:Cluster_analyses}. %

\begin{figure}[!htb]
    % \vspace{-10pt}
    \centering
    \includegraphics[width=1\linewidth]{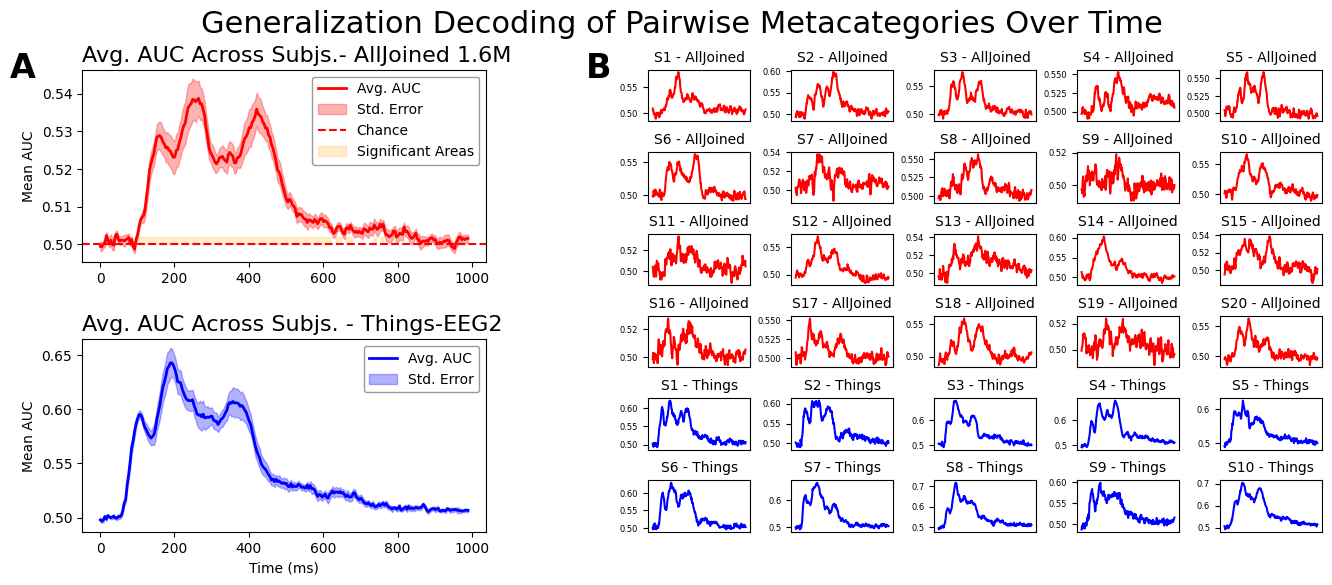}
    \vspace{-15pt}
    \caption{Average pair-wise decoding across meta-category combinations. Fig. \ref{fig:generalization}A depicts the average decoding AUC scores across all 20 Alljoined-1.6M subjects (red) and across all 10 THINGS-EEG2 subjects (blue). Results show significant decoding compared to baseline. Fig. \ref{fig:generalization}B Depicts the corresponding AUC performance scores for all individual subjects in both datasets.}
    \label{fig:generalization}
    \vspace{-15pt}
\end{figure}

\textbf{Pairwise Decoding.} To examine when category information becomes linearly separable, we first adopted a pairwise Linear Discriminant Analysis (LDA) decoder. LDA is intentionally simple: it is fast enough to evaluate every post-stimulus sample, needs no hyper-parameter tuning, and yields a single weight vector per class pair, making the decision boundary transparent. Using the training split, we fit LDA models and computed time-resolved ROC-AUC on held-out trials. A cluster-based permutation test against the 0.50 chance level (multiple-comparison corrected) revealed significant clusters (\textit{p} < 0.01) peaking around 100 ms, 220 ms, and 400 ms after stimulus onset (Fig. \ref{fig:generalization}). Although the consumer-grade Emotiv headset in Alljoined-1.6M introduces more noise than the research-grade hardware used in THINGS-EEG2, the decoder still achieved robust above-chance performance, replicating the temporal structure reported with high-quality systems. Thus, a lightweight, interpretable linear model is sufficient to expose the key temporal dynamics of category-level signals in this dataset while providing a clear baseline for more complex approaches.

\begin{figure}[!htb]
\centering
\vspace{-10pt}
\includegraphics[width=\textwidth]{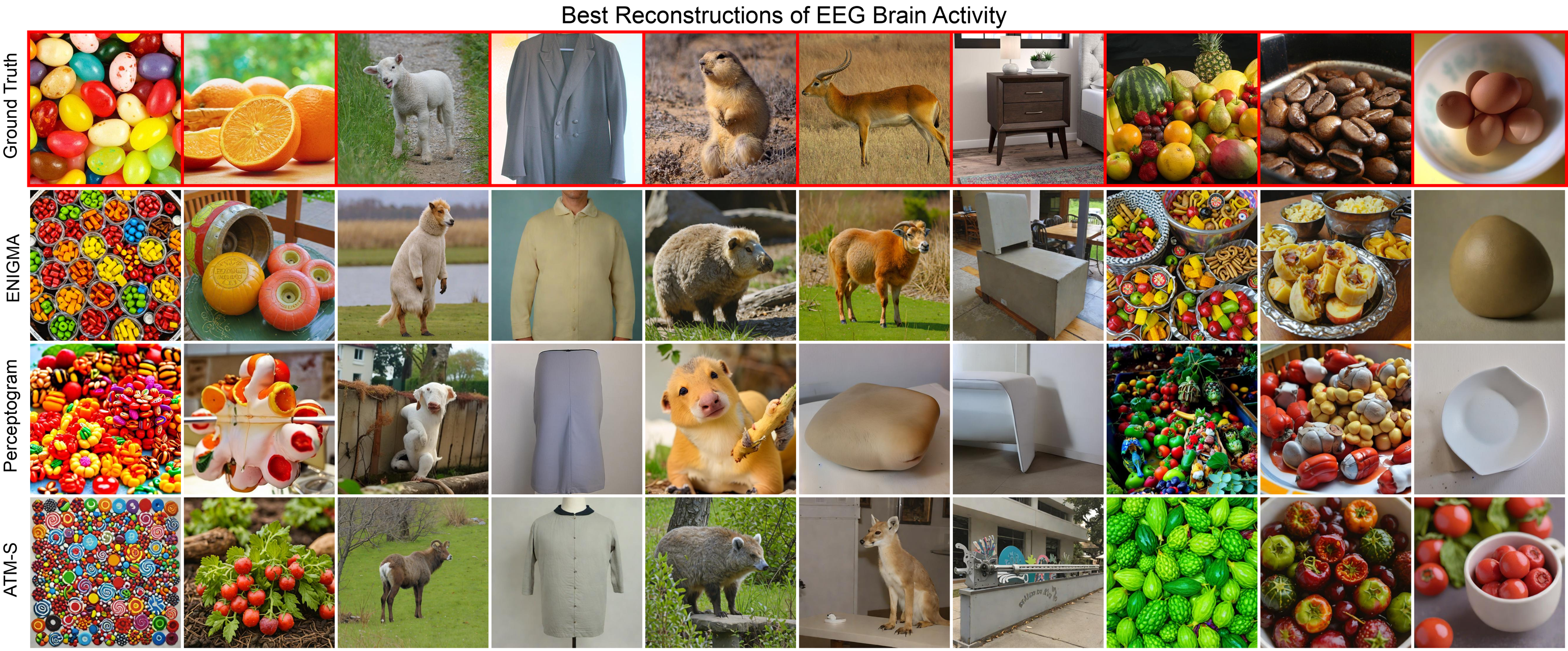}
\vspace{-5pt}
\caption{Qualitative comparison of EEG-to-Image reconstruction methods on Alljoined-1.6M. Reconstructions selected are the outputs sampled from each method and stimulus with the highest scores on all of the image feature metrics in Table \ref{table:featuremetrics}.}
\label{figure:best}
\vspace{-10pt}
\end{figure}

\textbf{EEG-to-Image Reconstruction.}
One of the most promising areas of BCI research is the development of models trained to reconstruct seen images from human brain activity. The number of research efforts tackling the adjacent task of fMRI-to-Image reconstruction tasks has taken off recently \cite{ozcelik2023braindiffuser, scotti_reconstructing_2023,Scotti2024MindEye2,takagi2022_decoding,takagi2023improving, kneeland_brain-optimized_2023, kneeland2023reconstructing, kneeland_second_2023, chen2023structure}. However EEG-to-Image efforts have lagged behind, with the first large-scale EEG-to-Image datasets THINGS-EEG1 and THINGS-EEG2 released only in 2022 \cite{Grootswagers2022, Gifford2022}. For our analysis, we took all publicly available EEG-to-Image reconstruction methods (ENIGMA \cite{ENIGMA}, ATM-S \cite{Li2024}, and Perceptogram \cite{Fei2024-perceptogram}) and reproduced their methods on our dataset. In line with research in fMRI-to-Image research \cite{Scotti2024MindEye2}, we also conducted a human behavioral experiment (n=$545$) to evaluate the identification accuracy of the reconstructions. Details on this behavioral experiment are provided in Appendix \ref{app:datadetails}. Image reconstructions from these methods can be seen in Fig. \ref{figure:best}, and quantitative results in Table \ref{table:featuremetrics}. We find that despite the high modeling difficulty of this task and the low SNR produced by the Emotiv hardware, several available EEG-to-Image reconstruction methods trained on Alljoined-1.6M produced reconstructions with quantitative scores comparable to those of THINGS-EE2 \cite{ENIGMA}. In our results we do notice that complex architectures like ATM-S \cite{Li2024} underperform on our data relative to simpler linear methods (Perceptogram \cite{Fei2024-perceptogram}, or more robust multi-subject models like ENIGMA \cite{ENIGMA}. We hope the release of Alljoined-1.6M will help spur further research translating existing approaches to the lower SNR data produced by the consumer-grade EEG setup in our study, as architecture clearly matters for bridging this gap.

\begin{table*}[!htb]
    \vspace{-5pt}
    \centering
    \setlength{\tabcolsep}{2pt}
    \small
    \resizebox{\textwidth}{!}{
    \begin{tabular}{lcccccccccccc}
        \toprule
        Method & \multicolumn{2}{c}{Low‑Level} & \multicolumn{6}{c}{High‑Level} & \multicolumn{3}{c}{Retrieval} & \multicolumn{1}{c}{Human Raters} \\
        \cmidrule(lr){2-3}\cmidrule(lr){4-9}\cmidrule(lr){10-12}\cmidrule(l){13-13}
        & PixCorr $\uparrow$ & SSIM $\uparrow$ & Alex(2) $\uparrow$ & Alex(5) $\uparrow$ & Incep $\uparrow$ & CLIP $\uparrow$ & Eff $\downarrow$ & SwAV $\downarrow$ & Top‑1 $\uparrow$ & Top‑5 $\uparrow$ & Top‑10 $\uparrow$ & Ident.\ Acc.\ $\uparrow$ \\
        \midrule
        \multicolumn{13}{c}{\textbf{THINGS‑EEG2}} \\
        \midrule
        ENIGMA & \underline{0.159} & \underline{0.422} & \underline{81.89\%} & \textbf{88.34\%} & \textbf{75.09\%} & \textbf{78.90\%} & \textbf{0.870} & \textbf{0.546} & \underline{27.60\%} & \underline{59.35\%} & \underline{71.15\%} & \textbf{83.06\%} \\
        ATM‑S           & 0.136 & 0.392 & 73.85\% & 80.83\% & 67.56\% & 71.28\% & 0.909 & 0.601 & \textbf{30.15\%} & \textbf{60.15\%} & \textbf{73.60\%} & 77.14 \\
        Perceptogram    & \textbf{0.247} & \textbf{0.431} & \textbf{85.46\%} & \underline{88.03\%} & \underline{70.40\%} & \underline{71.98\%} & \underline{0.902} & \underline{0.581} & -- & -- & -- & \underline{79.17\%} \\
        \midrule
        \multicolumn{13}{c}{\textbf{Alljoined‑1.6M}} \\
        \midrule
        ENIGMA & 0.079 & \textbf{0.416} & \underline{63.62\%} & \underline{67.84\%} & \textbf{59.57\%} & \textbf{62.91\%} & \textbf{0.942} & \textbf{0.620} & \textbf{6.00\%} & \textbf{16.25\%} & \textbf{25.35\%} & \textbf{65.43\%} \\
        ATM‑S          & \underline{0.090} & 0.374 & 55.91\% & 58.25\% & 54.07\% & 56.25\% & 0.960 & 0.673 & \underline{0.50\%} & \underline{2.00\%} & \underline{5.00\%} & 60.31\% \\
        Perceptogram    & \textbf{0.094} & \underline{0.401} & \textbf{67.36\%} & \textbf{69.28\%} & \underline{58.18\%} & \underline{59.94\%} & \underline{0.945} & \underline{0.637} & -- & -- & -- & \underline{62.00\%} \\
        \bottomrule
    \end{tabular}
    }
    \caption{Quantitative comparison between reconstruction quality of available methods on the THINGS‑EEG2 and Alljoined‑1.6M datasets. PixCorr is the pixel-level correlation score. SSIM is the structural similarity index metric \cite{wang_image_2004}. AlexNet($2$) and AlexNet($5$) are the 2-way comparisons (2WC) of layers 2 and 5 of AlexNet \cite{alexnet}. CLIP is the 2WC of the output layer of the CLIP ViT-L/14 Vision model \cite{radford2021learning}. Incep is the 2WC of the last pooling layer of InceptionV3 \cite{inceptionv3}. EffNet-B and SwAV are distance metrics gathered from EfficientNet-B13 \cite{tan_efficientnet_2020} and SwAV-ResNet50 \cite{caron_unsupervised_2021} models. Details on the human identification accuracy metric are provided in Appendix \ref{app:behavioral}. For EffNet-B and SwAV distances, lower is better. For all other metrics, higher is better. Bold indicates best performance, and underlines second-best performance. Additional details on the metrics used are in Appendix \ref{app:datadetails}.} %\ref{app:metrics}
    \vspace{-8pt}
    \label{table:featuremetrics}
\end{table*}

\begin{figure}
    \centering
    \includegraphics[width=\linewidth]{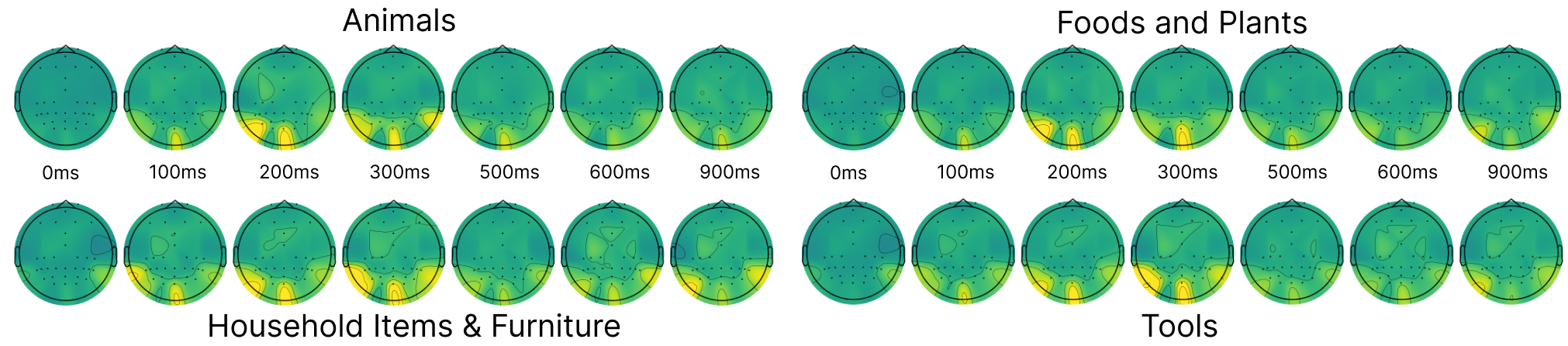}
    \vspace{-5pt}
    \caption{\textbf{Spatiotemporal saliency for ENIGMA.} 
    Warm colors indicate electrodes whose activity diverges most strongly from the grand-average VEP and therefore pushes the model toward the CLIP centroid of that category. All four rows reveal a shared peak over occipital sensors around 160–300 ms (yellow), consistent with the early P1/N1 complex that dominates low-level visual processing.}
    \label{fig:saliency}
    \vspace{-5pt}
\end{figure}

\textbf{Saliency Maps.} 
To locate the spatiotemporal features that underpin our semantic predictions, we applied Integrated Gradients \citep{sundararajan2017axiomatic} with each meta-category’s mean CLIP embedding as the target. For every category (animals, household items, foods/plants, tools) we averaged the training-image CLIP vectors, computed attributions over the full EEG tensor (channels × time), smoothed them with an 11-sample boxcar, and projected the result onto the 10-20 montage. All categories yield a pronounced attribution peak over occipital sensors at 160–300 ms post-stimulus (Fig. \ref{fig:saliency}), pointing to early visual activity \citep{cichy2014resolving, thorpe1996speed}.

Because our baseline is the grand-average EEG, the maps capture deviations from the canonical VEP that the network uses to match each category’s CLIP centroid. The virtually identical occipital P1/N1 footprint across categories implies that the model relies almost entirely on low-level visual cues—an outcome we ascribe partly to the constraints of the RSVP paradigm \citep{holm2023contribution}.

\begin{figure}[!htb]
\vspace{-5pt}
  \centering
  %—— minipage A: Exp4 plot ——
  \begin{minipage}[t]{0.49\textwidth}
    \centering
    \textbf{A}\\[4pt]
    \includegraphics[width=\linewidth]{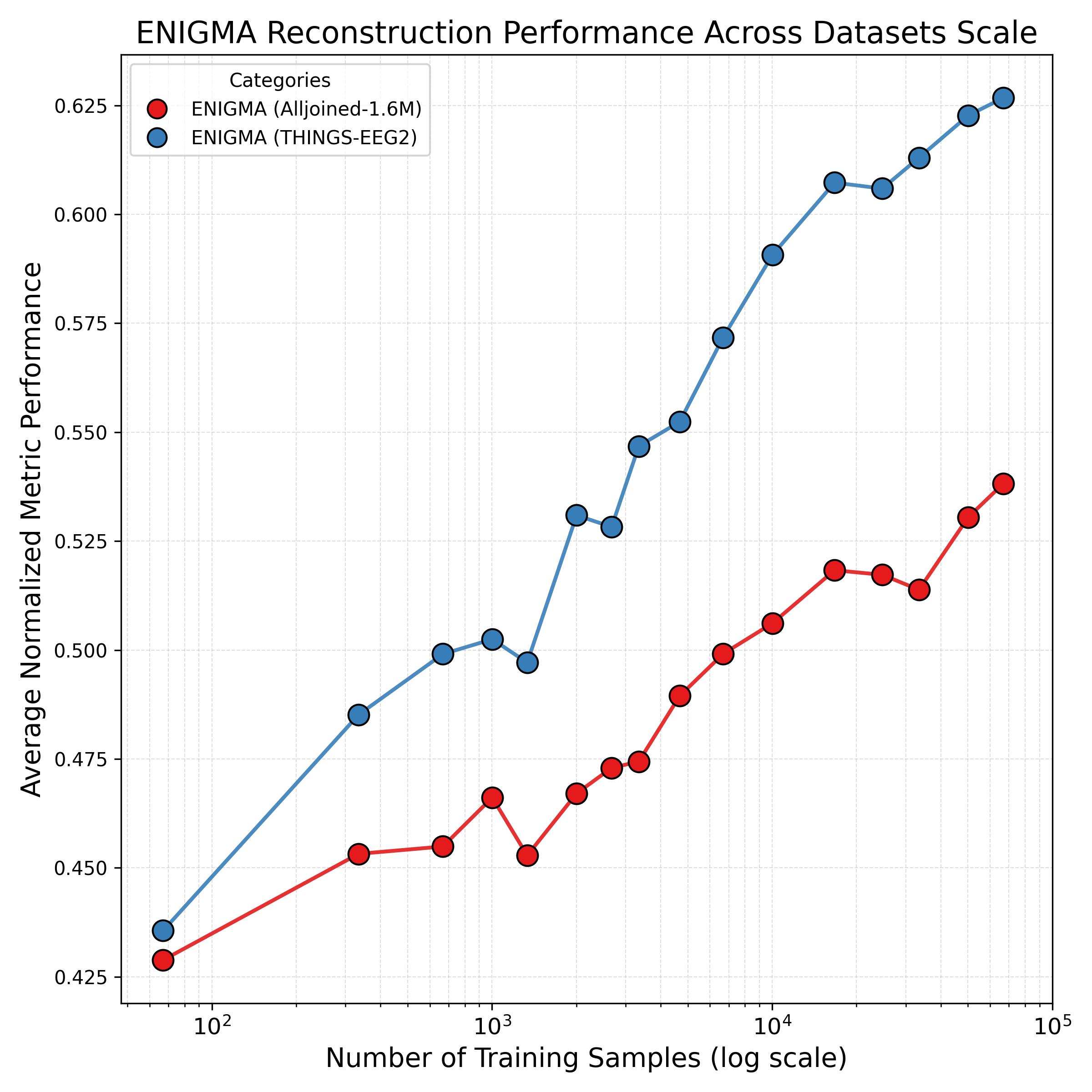}
  \end{minipage}\hfill
  %—— minipage B: Ablation plot ——
  \begin{minipage}[t]{0.49\textwidth}
    \centering
    \textbf{B}\\[4pt]
    \includegraphics[width=\linewidth]{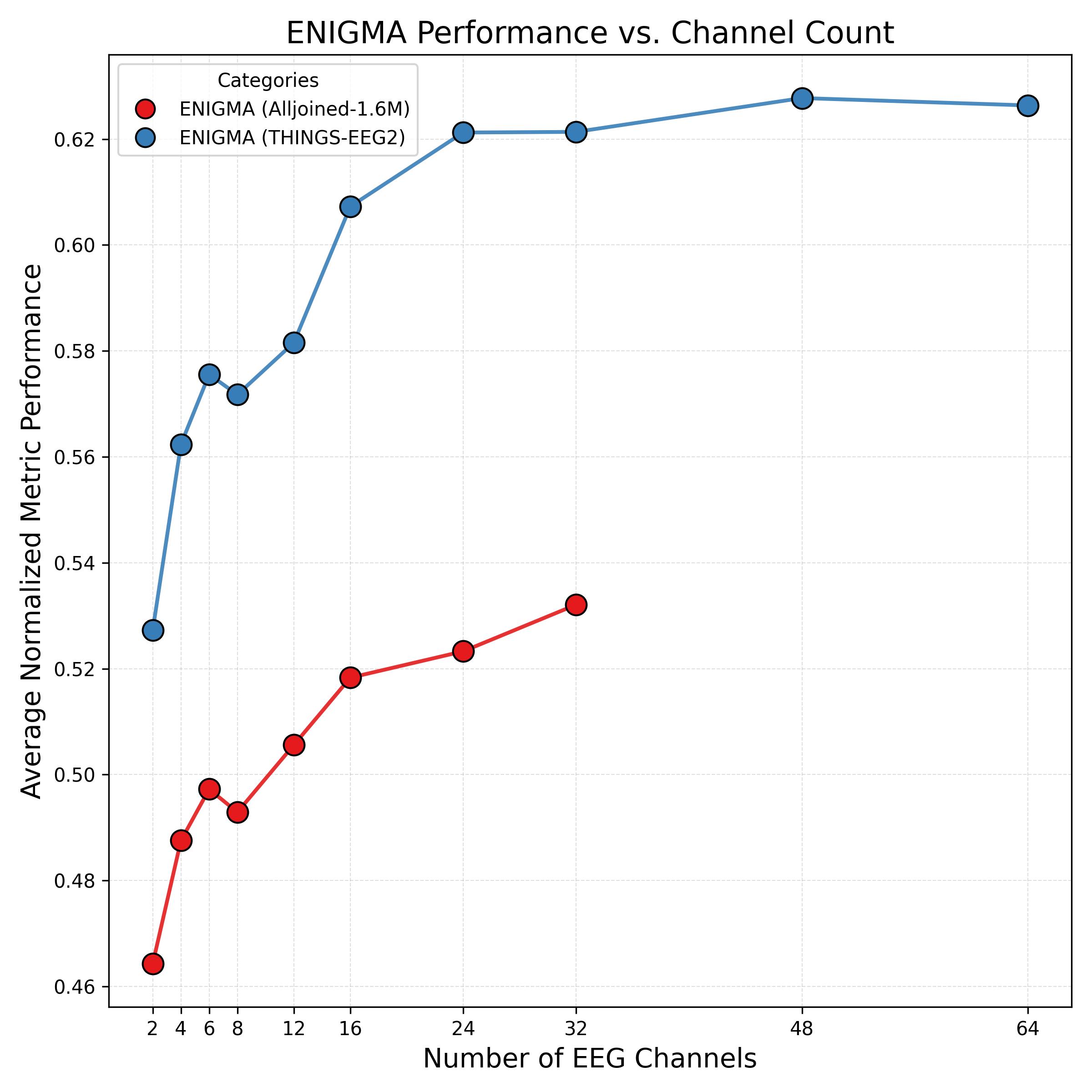}
  \end{minipage}

  \vspace{-6pt}
  \caption{%
    \textbf{(A)} Scaling analysis of model performance for Alljoined-1.6M and THINGS-EEG2. The number of training samples are plotted on a log-scale X-axis, and the normalized average of feature metrics presented in Table \ref{table:featuremetrics} is plotted on the Y-axis.
    \textbf{(B)} Channel count analysis of model performance for each dataset. The number of channels in each dataset was progressively reduced, while the remaining channels focus primarily on occipital cortex. The Y axis is plotted the same as Fig. \ref{fig:scaling}A.
  }
  \vspace{-8pt}
  \label{fig:scaling}
\end{figure}

\textbf{Scaling Analysis.} To contextualize Alljoined‑1.6M within current scaling debates, we followed Banville et al.’s subsampling protocol \cite{banville2025scaling} and trained ENIGMA \cite{ENIGMA}, the leading EEG‑to‑Image model, on progressively larger subsets. Reconstruction quality was summarized by the normalized mean of the metrics in Table \ref{table:featuremetrics}. Figure \ref{fig:scaling}A plots ENIGMA’s log–log learning curves for Alljoined‑1.6M and, for comparison, the higher‑SNR THINGS‑EEG2. Performance increased almost linearly with log‑trial count and showed no sign of saturating at the full dataset size. As expected, the consumer‑grade recordings scaled less efficiently, underscoring the noise penalty of low‑cost headsets. This limitation is also a strength: our dataset provides a realistic benchmark for methods that must cope with low‑SNR data. The pattern mirrors findings across machine learning: more data reliably boosts accuracy - and the low price of consumer hardware should allow still larger datasets in the future, potentially offsetting the SNR gap through sheer quantity.

\textbf{Channel Count Analysis.} One of the most obvious differences between the research grade ActiChamp amplifier used in the THINGS-EEG2 dataset and the Emotiv Flex 2 hardware used in our dataset is the number of channels (64 vs 32). We performed an analysis to evaluate how the number of channels affects decoding performance using the ENIGMA model, and to explore whether this difference was a significant contributor to differences in performance. We sub-sampled both datasets to varying numbers of electrodes, while retaining a focus on covering occipital cortex with the electrodes selected. We find that while performance did drop with fewer channels, it is not the most significant factor accounting for the performance difference between the datasets, and that performance gains start to drop off after 24 channels, suggesting that future studies might still be able to achieve reasonable decoding performance with even fewer channels.

\section{Discussion}
\label{sec:discussion}
We have introduced \textbf{Alljoined-1.6M}, the largest publicly available EEG dataset for visual cognition to date, and collected using a consumer-grade EEG headset. Our analyses provide encouraging evidence that EEG data collected on consumer-grade EEG hardware—such as the Emotiv Flex 2 utilized in our study—is still rich enough to train modern semantic decoding algorithms, a promising sign for the development of affordable brain-computer interfaces! We also find that scaling up data collection remains an effective way of increasing decoding performance even despite hardware limitations, opening doors to new large scale data collection efforts for affordable hardware. These findings have significant implications for the future of BCI research and cognitive neuroscience: large‑scale EEG acquisition is now feasible for small labs, classrooms, and citizen‑science projects. Alljoined‑1.6M therefore serves as a benchmark for algorithmic progress, a blueprint for affordable data collection, and a concrete step toward democratizing neurotechnology.

\textbf{Broader Impacts.}
\label{broaderimpacts}
Alljoined-1.6M highlights many considerations for the design of future datasets in brain decoding. While many efforts to collect neuroimaging datasets emphasize high channel counts or ultra-high-resolution signals, we believe that we need more datasets that are representative of real-world use cases. Our results (among others \cite{banville2025scaling}) also point to the underexplored value of sheer volume, repetition, and participant diversity—factors that become significantly more tractable with affordable hardware. Alljoined-1.6M is a promising first step in shifting away from collecting a small amount of pristine signal, to instead optimizing for scale, signal diversity, and accessibility. We suggest that future datasets prioritize these axes, leveraging low-cost, high-throughput paradigms to explore larger-scale representational learning across subjects and tasks, much like large vision or language datasets \citep{devlin2019bert, brown2020language, deng2009imagenet} have done for deep learning. We envision a future where brain data collection is not bottle-necked by cost, and where massive EEG datasets fuel breakthroughs in understanding the brain and building BCI technologies to make a difference to people around the world.

\textbf{Limitations and Future Work.}
\label{limitations}
Our dataset evaluates only one low-cost consumer-grade headset (Emotiv Flex 2). Testing other affordable devices, and guiding research-based product design in amplifiers and materials could further boost signal quality and sharpen the cost–performance frontier. We also observed roughly log-linear scaling; rigorously tracking accuracy as we grow from $10^6$ to $10^7$ trials, and exploring smarter sampling or augmentation strategies should help to further clarify these dynamics. Because the current corpus involves healthy adults in a controlled lab, future efforts should gather at-home, asynchronous recordings from diverse populations, transforming EEG collection into crowd-sourced neuroscience. It may also be exciting to merge multiple low-cost wearables or hybrid EEG + peripheral sensor arrays to narrow the gap to clinical-grade rigs. Methodologically, we observe the RSVP paradigm drops SNR at later latencies, so alternative task designs warrant exploration. Finally, our release offers a testbed for large-scale, multi-subject modeling: training a single network on the full corpus could yield generalizable neural representations transferable across users and downstream tasks, paralleling recent self-supervised EEG work \citep{Banville2021}.

\textbf{Ethical Considerations.}
\label{ethics}
Efforts to collect and utilize neuroimaging datasets of human brain activity is rapidly growing in scale and capability. While this research promises clear downstream benefits in a variety of applications, we believe it is important to consider the ethical burden of gaining access to the internal cognitive states of individuals, and we recognize the potential for this technology to be misused. It is therefore important to begin developing an ethical framework for the application of brain-decoding devices and datasets that rigorously safeguards users data \cite{sethics}, and ensures that the technology is deployed transparently, responsibly, and for the benefit of humankind.

\bibliographystyle{plain}
\bibliography{main}

\newpage
\appendix
\input{appendix}

\end{document}

%% file: appendix.tex
\section{Appendix}

\subsection{Demographics}
\label{app:demographics}
\begin{table}[htbp]
\caption{\textit{Participant Demographics by Category}}
\centering
\begin{tabular}{llll}
\toprule
\textbf{Variable} & \textbf{Category} & \textbf{Count} & \textbf{Percentage} \\
\midrule
Sex & Male & 15 & 75 \\
 & Female & 5 & 25 \\
 & Other & 0 & 0 \\
Age & 18--25 & 3 & 15 \\
 & 26--35 & 3 & 15 \\
 & 36--45 & 6 & 30 \\
 & 46--55 & 7 & 35 \\
 & 55+ & 1 & 5 \\
Ethnicity & White & 4 & 20 \\
 & Black or African American & 1 & 5 \\
 & Asian & 3 & 15 \\
 & Hispanic/Latino & 4 & 20 \\
 & Native Hawaiian or Pacific Islander & 2 & 10 \\
 & American Indian or Alaska Native & 1 & 5 \\
 & Multiracial & 0 & 0 \\
 & Declined to answer & 5 & 25 \\
Handedness & Right & 17 & 85 \\
 & Left & 0 & 0 \\
 & Ambidextrous & 3 & 15 \\
Recruitment Platform & Craigslist & 11 & 55 \\
 & Instawork & 7 & 35 \\
 & Other & 2 & 10 \\
\bottomrule
\end{tabular}
\end{table}

\subsection{Data collection setup}
\label{app:setup}
\begin{figure}[!htb]
\centering
\includegraphics[width=0.8\textwidth]{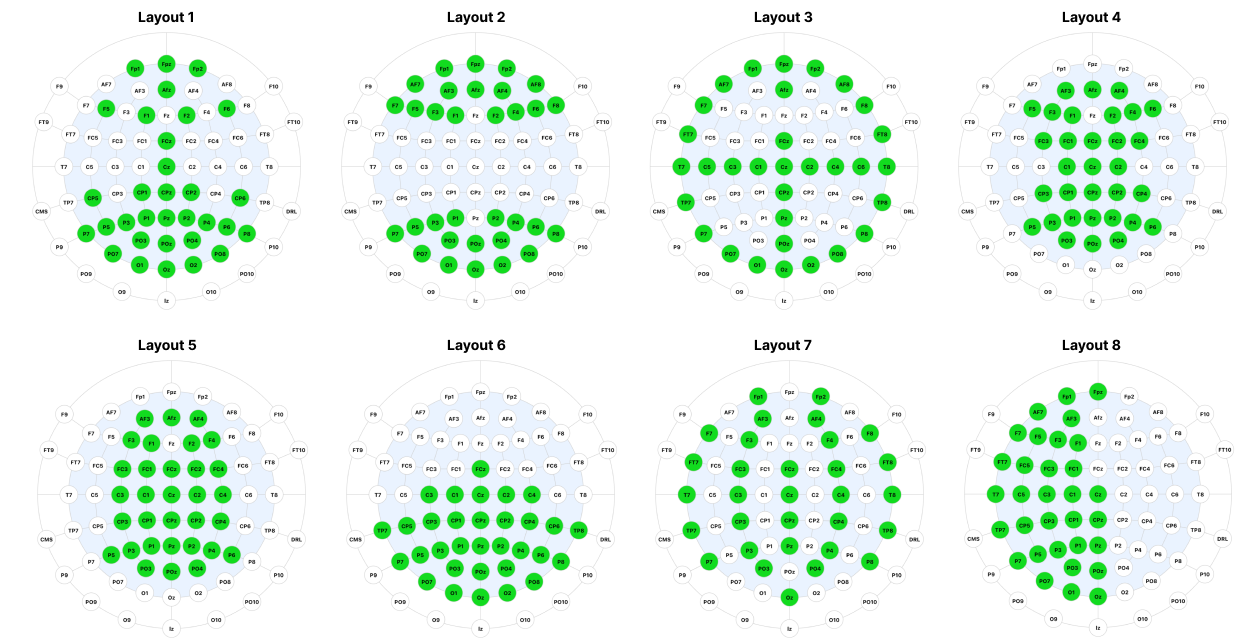}
\vspace{-5pt}
\caption{Various electrode positions and subsets from the 10-20 layout, compared in ablations on downstream decoding performance in Figure \ref{figure:position_ablations}.}
\label{fig:electrodes}
\end{figure}

The electrode positions selected for data collection followed the standard 10-20 format, and were focused on the occipital region and central line  (Cz, Fp1, F7, F3, CP5, CP1, P1, P3, P5, P7, PO9, PO7, PO3, O1, O9, Pz, POz, Oz, O10, O2, PO4, PO8, PO10, P8, P6, P4, P2, CP2, CP6, F4, F8, Fp2). This corresponds to Layout 1 in \ref{fig:electrodes}). These were selected by running ablations of decoding performance on the Things-EEG2 dataset.

\begin{figure}[!htb]
\centering
\includegraphics[width=0.6\textwidth]{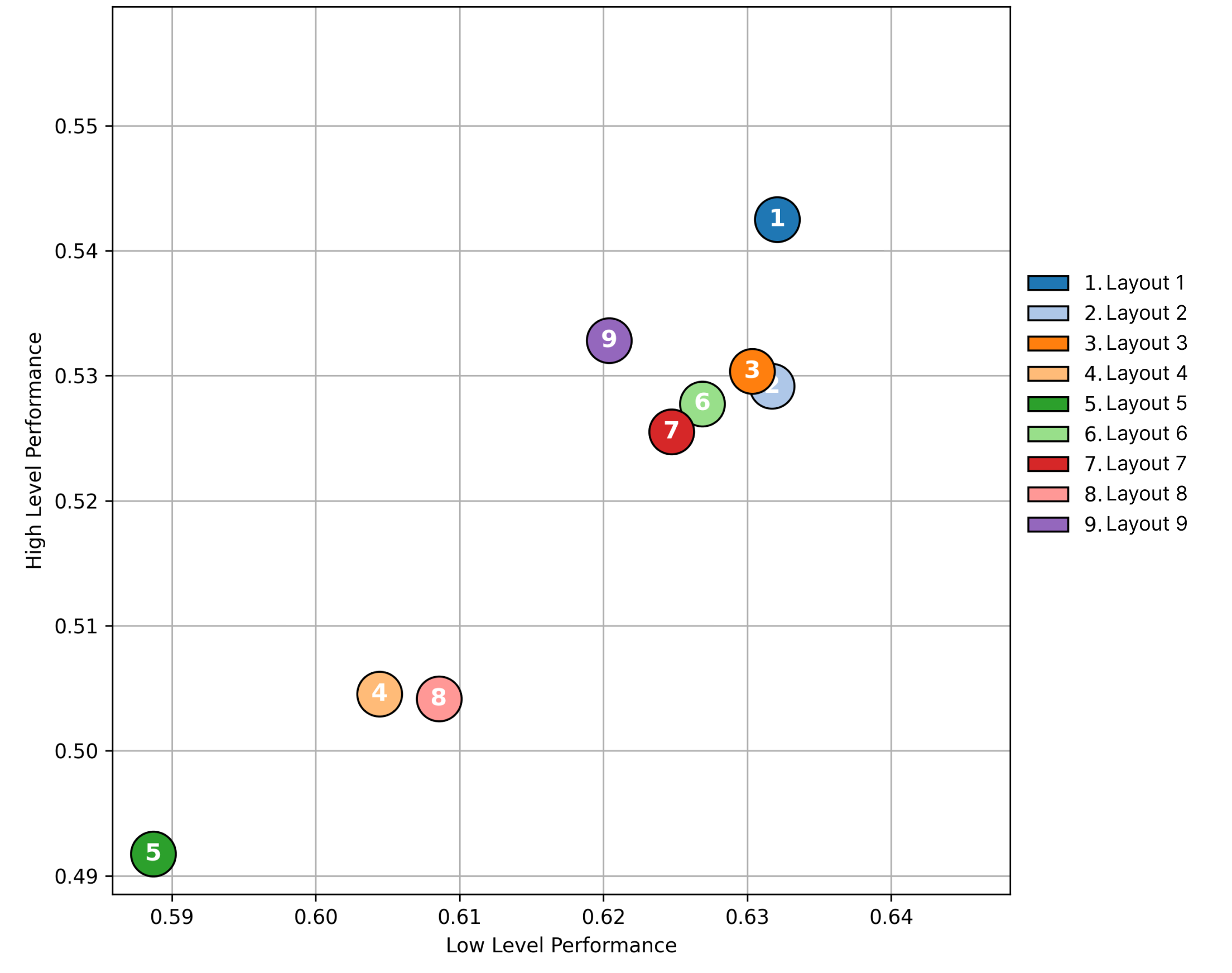}
\vspace{-5pt}
\caption{High and low level performance for image reconstruction on various 32 channel electrode subsets of THINGS-EEG2. Axes are the normalized average of the metrics in the high and low level metric categories displayed in Table \ref{table:featuremetrics}.}
\label{figure:position_ablations}
\end{figure}

\subsection{Data Collection Details}
\label{app:datadetails}
Of the 48 participants initially recruited for this study, a total of 28 were subsequently excluded from the analyses for the following reasons. First, 18 participants were unable to commit to the three subsequent data collection sessions due to personal constraints, e.g., scheduling conflicts, illness, or travel. Second, despite our best efforts to coach and stabilize them, 6 participants produced data of insufficient quality as they were not able to sit still long enough for reliable EEG acquisition, resulting in excessive noise or artifacts. Finally, 4 participants were removed on the basis of behavioral and interpersonal factors: they were difficult or unpleasant to work with, or were unable to follow instructions for the experiment. The remaining 20 participants comprised the final dataset reported in this paper.

\subsection{Behavioral Performance Analysis}
\label{app:behavioral_perf}
\begin{figure}[h!]
    \centering
    \includegraphics[width=\textwidth]{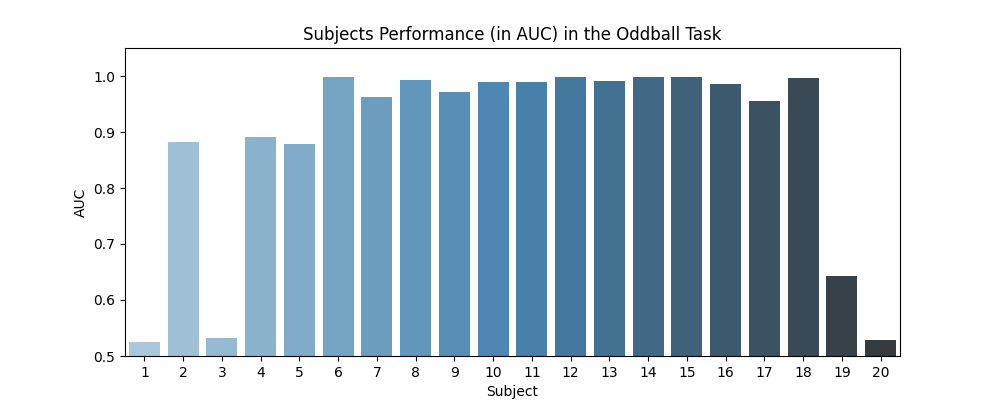}
    \caption{AUC values for each of the 20 participants based on their behavioral responses in the RSVP task. Higher AUC indicate better discrimination between target-present and target-absent trials, independent of response bias.}
    \label{fig:auc_appendix}
\end{figure}

During the experiment, we conducted regular attention checks in a manner very similar to the attention checks reported in the THINGS-EEG2 dataset \citep{Gifford2022}. Subjects were given the "odd-ball" task of looking for a picture of Woody from the Toy Story movies, and were asked at the end of each block whether an image of Woody appeared. We collected the accuracy of subjects on this detection task. However, since approximately 94\% of the RSVP blocks did not contain the target (Woody), accuracy scores would be inflated by a strong response bias toward saying "no." That is, even random guessing would yield a high accuracy. To correct for this bias and better assess true sensitivity to the presence of the target, we computed the area under the receiver operating characteristic curve (AUC) for each participant and plotted them in Figure \ref{fig:auc_appendix}.

Subjects performed at a 88\% AUC in the task, with a standard error of 1\%. Three subjects performed close to chance rates, but all other subjects performed close to 100\%. These results demonstrate that most subjects were attentive to the task.

The AUC reflects the probability that a randomly chosen trial with the target present is rated as more likely to contain the target than a randomly chosen target-absent trial. It is robust to response bias and provides a more balanced measure of detection performance in imbalanced-class settings. An AUC of 0.5 indicates chance-level performance, whereas 1.0 indicates perfect sensitivity.

We also note in our analysis that the three subjects that performed poorly on the attention task still yielded strong decoding performance on the visual decoding experiments, and so we keep these subjects in our dataset.

\FloatBarrier
 
\subsection{Additional Details on Evaluation Metrics}
\label{app:metrics}
We use the following image similarity metrics:
\begin{itemize}
\item PixCorr is the pixel-level correlation between the ground-truth images and reconstructed images.
\item SSIM is the structural similarity index metric \cite{wang_image_2004}.
\item AlexNet($2$) and AlexNet($5$) are the 2-way comparisons (2WC) of layers 2 and 5 of AlexNet \cite{alexnet}.
\item CLIP is the 2WC of the output layer of the CLIP ViT-L/14 Vision model \cite{radford2021learning}.
\item Incep is the 2WC of the last pooling layer of InceptionV3 \cite{inceptionv3}.
\item Eff and SwAV are distance metrics gathered from EfficientNet-B13 \cite{tan_efficientnet_2020} and SwAV-ResNet50 \cite{caron_unsupervised_2021} models. 
\end{itemize}
For the metrics in Table $1$, a two-way comparison (2WC) evaluates whether the feature embedding of the stimulus image is more similar to the feature embedding of the target reconstruction, or the feature embedding of a randomly selected "distractor" reconstruction, where the score is the percent of correctly identified target reconstructions across a pool of "distractors". Our 2WC metrics, calculated using reconstructions of the $199$ other test-set stimuli as "distractors", have a notably different chance threshold from 2WC metrics presented in reconstruction papers that perform evaluations using a test set with a different number of "distractors", such as the shared1000 test set of NSD \cite{allen_massive_2022}, and are thus not directly comparable. All metrics in Table 1 were calculated and averaged across 10 images sampled from the output distribution of each method using a random seed. All metrics in Table were calculated on our reproduction of other methods using their open source code, and might differ slightly from metrics reported in the original papers due to our implementation of the metrics we calculated.

\FloatBarrier
\clearpage
\subsection{Statistical Significance of Metrics}
\label{app:statisticalsig}

\begin{table*}[!htb]
    \vspace{-10pt}
    \centering
    \setlength{\tabcolsep}{2pt}
    \small
    \resizebox{\textwidth}{!}{
    \begin{tabular}{lccccccccc}
        \toprule
        Method & \multicolumn{4}{c}{Low-Level} & \multicolumn{4}{c}{High-Level} & \multicolumn{1}{c}{Human Raters} \\
        \cmidrule(lr){2-5}\cmidrule(lr){6-9}\cmidrule(l){10-10}
        & PixCorr $\uparrow$ & SSIM $\uparrow$ & Alex(2) $\uparrow$ & Alex(5) $\uparrow$ & Incep $\uparrow$ & CLIP $\uparrow$ & Eff $\downarrow$ & SwAV $\downarrow$ & Ident.\ Acc.\ $\uparrow$ \\
        \midrule
        \multicolumn{10}{c}{\textbf{THINGS-EEG2}} \\
        \midrule
        ENIGMA (multi-subject) & $\pm$0.0014 & $\pm$0.0014 & $\pm$0.15\% & $\pm$0.12\% & $\pm$0.20\% & $\pm$0.19\% & $\pm$0.0008 & $\pm$0.0008 & $\pm$0.89\% \\ 
        ATM-S (multi-subject) & $\pm$0.0009 & $\pm$0.0010 & $\pm$0.20\% & $\pm$0.20\% & $\pm$0.21\% & $\pm$0.21\% & $\pm$0.0004 & $\pm$0.0006 & $\pm$1.15\% \\ 
        \midrule
        ENIGMA (single-subject) & $\pm$0.0014 & $\pm$0.0014 & $\pm$0.14\% & $\pm$0.11\% & $\pm$0.20\% & $\pm$0.18\% & $\pm$0.0008 & $\pm$0.0008 & $\pm$0.87\% \\ 
        ATM-S (single-subject) & $\pm$0.0013 & $\pm$0.0013 & $\pm$0.17\% & $\pm$0.15\% & $\pm$0.21\% & $\pm$0.20\% & $\pm$0.0007 & $\pm$0.0008 & $\pm$0.97\% \\ 
        Perceptogram (single-subject) & $\pm$0.0014 & $\pm$0.0015 & $\pm$0.12\% & $\pm$0.11\% & $\pm$0.20\% & $\pm$0.20\% & $\pm$0.0007 & $\pm$0.0007 & $\pm$0.94\% \\ 
        \midrule
        \multicolumn{10}{c}{\textbf{Alljoined-1.6M}} \\
        \midrule
        ENIGMA (multi-subject) & $\pm$0.0007 & $\pm$0.0008 & $\pm$0.14\% & $\pm$0.13\% & $\pm$0.15\% & $\pm$0.15\% & $\pm$0.0005 & $\pm$0.0005 & $\pm$0.77\% \\ 
        ATM-S (multi-subject) & $\pm$0.0007 & $\pm$0.0007 & $\pm$0.14\% & $\pm$0.15\% & $\pm$0.15\% & $\pm$0.15\% & $\pm$0.0003 & $\pm$0.0004 & $\pm$0.82\% \\ 
        \midrule
        ENIGMA (single-subject) & $\pm$0.0007 & $\pm$0.0009 & $\pm$0.14\% & $\pm$0.14\% & $\pm$0.15\% & $\pm$0.15\% & $\pm$0.0004 & $\pm$0.0005 & $\pm$0.78\% \\ 
        ATM-S (single-subject) & $\pm$0.0007 & $\pm$0.0008 & $\pm$0.14\% & $\pm$0.15\% & $\pm$0.15\% & $\pm$0.15\% & $\pm$0.0004 & $\pm$0.0005 & $\pm$0.80\% \\ 
        Perceptogram (single-subject) & $\pm$0.0008 & $\pm$0.0010 & $\pm$0.13\% & $\pm$0.13\% & $\pm$0.15\% & $\pm$0.15\% & $\pm$0.0004 & $\pm$0.0005 & $\pm$0.79\% \\ 
        \bottomrule
    \end{tabular}

    }
    \vspace{-5pt}
    \caption{Standard error measurements for evaluation metrics of EEG-to-Image reconstruction models evaluated on the THINGS‑EEG2 and Alljoined-1.6M datasets. Values correspond to the standard error spread of values in Table \ref{table:featuremetrics} in the manuscript.} %\ref{app:metrics}
    \label{table:statisticalsignificance}
\end{table*}
\FloatBarrier

\subsection{Behavioral Evaluation Experiments}
\label{app:behavioral}
To evaluate the quality of EEG-to-Image reconstruction models applied to our dataset, we conducted a behavioral experiment on $545$ human raters online. For our experiment, we identified no risks to the human participants, and collected informed consent from all participants. 

The experiment stimuli consists of image reconstruction sampled from the 30 subjects across THINGS-EEG2 and Alljoined-1.6M from all methods and cases in Table \ref{table:featuremetrics}. The images were shuffled and $60$ images presented to each subject. We use attention checks to identify whether human raters were paying attention to the task and the instructions and dropped $8$ human raters who failed at least $2$ out $8$  attentions checks before analysis. An attention check presents the ground truth image as one of the candidate images and raters have to select the candidate ground truth image (as an image is most similar to itself) to pass.

Our subjects were recruited through the \href{http://www.prolific.ac.uk)}{Prolific platform}, with our experimental tasks hosted on \href{http://meadows-research.com}{Meadows}. Each human rater was paid $\$1.25$ for the completion of the experiment, and the median completion time was $5$ minutes, resulting in an average payment rate of $\$15$/hour. The code to reproduce our experiment can be found in \href{https://github.com/Alljoined/Alljoined-1.6M}{our GitHub repository.}

\subsubsection{2AFC identification task}
\begin{figure}[!htb]
\setlength{\belowcaptionskip}{-10pt}    % Space below the caption
\begin{center}
\includegraphics[width=\columnwidth]{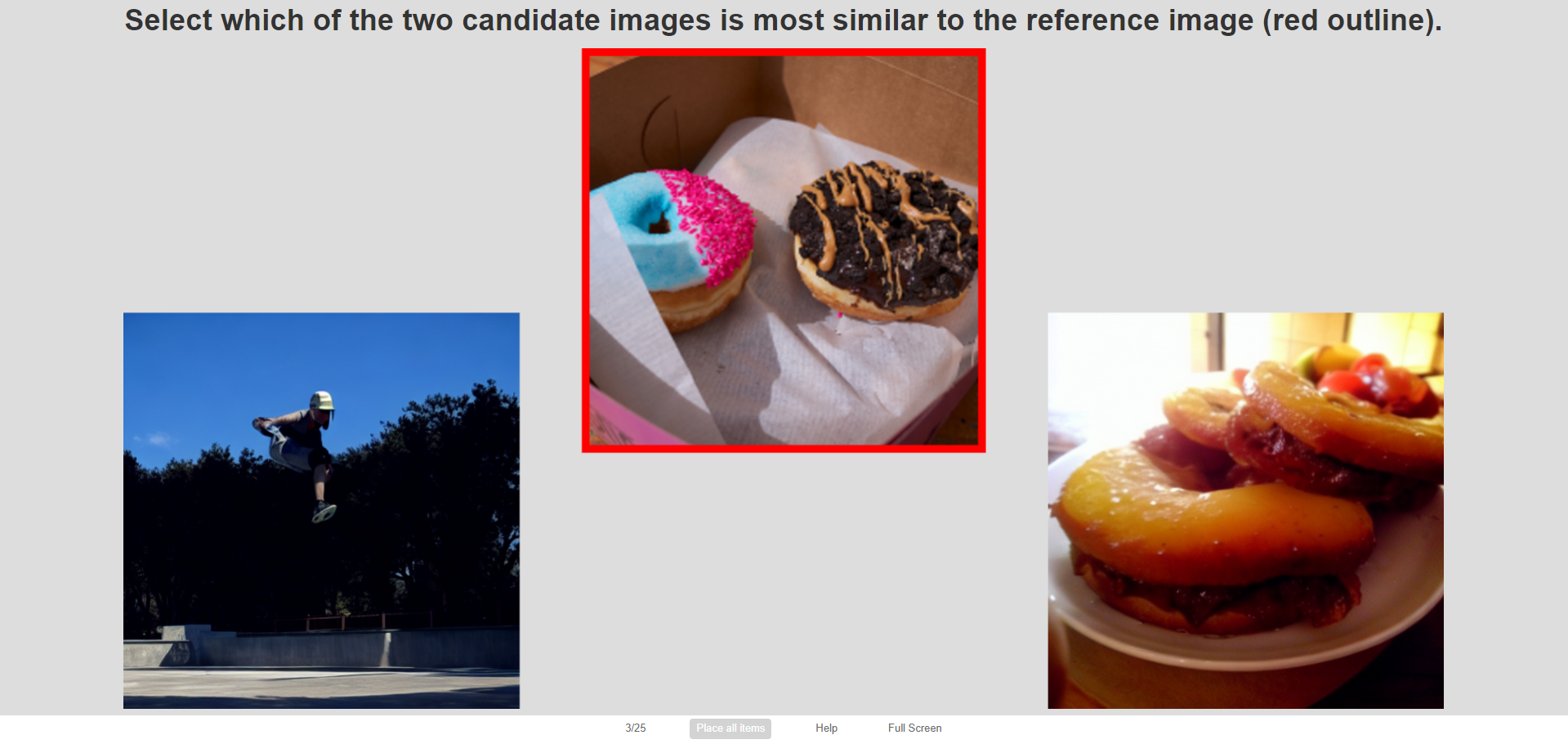}
\end{center}
\caption{An example of the 2 alternative forced choice task used in our behavioral experiment performed by human raters.} 
\label{figure:task1}
\end{figure}

Our experiment was a $2$ alternative forced choice task (2AFC) facilitated by the "Match-To-Sample" task on the Meadows platform. An example of the first experiment can be seen in Figure \ref{figure:task1}. In this experiment, human raters were asked to select which of two candidate images was more similar to a reference image. The reference image provided is the ground truth image the subject either saw, and the $2$ candidate images were the target reconstruction of the reference image, or a randomly selected reconstruction from an EEG recording corresponding to a different stimulus. The two candidate images were always sampled from the same reconstruction method and subject. This experiment was repeated for all reconstruction methods, model types, datasets, and subjects. With the results presented in Table \ref{table:featuremetrics}, we establish a baseline for human-rated image identification accuracy of seen image reconstructions from EEG, as no other paper has conducted behavioral evaluations of EEG-to-Image reconstructions.

\subsection{Meta-Categories}
\label{app:supercategories}
To create the meta-categories, we first used ChatGPT 4o and Gemini 2.5 Pro Preview to organize the original 1854 categories into groups. Note that the above GenAI tools often skipped words, misspelled them, miscategorized them, and sometimes hallucinated new words altogether. Therefore, we started with AI-generated categories and manually organized the images categories into meta-categories by also checking ambiguous or confusing categories (e.g., "mullet" refers to hair and not to fish). After visual inspection, we noticed that the test set did not contain any buildings and outdoor scenes, and excluded them from our analysis. We also noticed that the categories musical instruments, and toys and games were much smaller than the other categories, so we grouped them together in a "Fun/Entertainment" Category, although we note that these categories potentially evoke different brain responses.

\subsection{Cluster Analyses}
\label{app:Cluster_analyses}
We ran all 21 contrasts between all pairs of metacategories, of which 16 had significant clusters. 
Most categories showed effects between 100 and 400ms, with the strongest and most consistent clusters emerging around 200ms post-stimulus at occipital electrodes—timing typically associated with early visual categorization components such as N170 and N200 \citep{bentin1996electrophysiological, eimer2000face}. Additional clusters around 400ms may reflect later event related potential (ERP) components such as N2 \citep{Polich2007} or N400 \citep{Kutas1980}, or possibly contrastive or integrative responses to successive images. Given the 200ms inter-stimulus interval in our RSVP design, neural responses to consecutive images are likely to overlap in time, leading to temporally smeared effects across stimuli \citep{fernandez2022effect, Cichy2014}. This overlap might delay or attenuate the emergence of late components like P300 (which we did not observe in subjects' ERPs \ref{fig:ERP}). Note that the data has been whitened, so the channels have no unit and do not look like traditional ERPs. 
\begin{figure}[!htb]
    \centering
    \includegraphics[width=\linewidth]{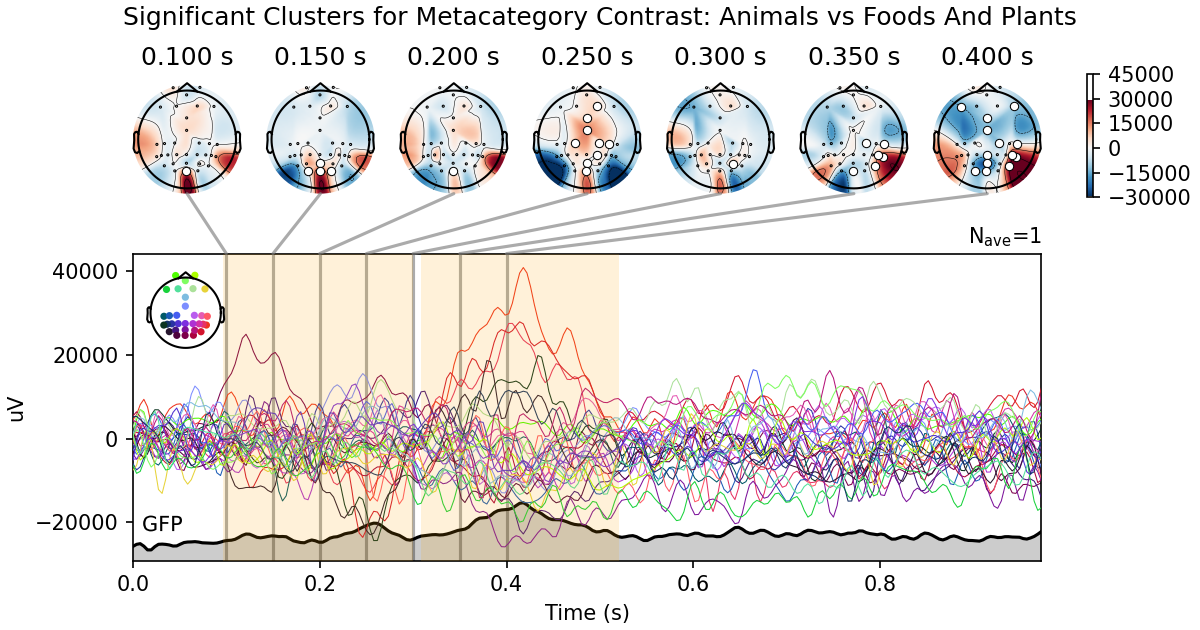}
    \caption{Cluster analyses results for the contrast between Animals and Foods/Plants. Yellow shaded area show times where difference between signals was significant. White dots in topographical maps represent significant electrodes which were significantly different for those time periods. Gray shaded area represents change in Global Field Power (GFP) over time. }
    \label{fig:cluster2}
\end{figure}

\begin{figure}[!htb]
    \centering
    \includegraphics[width=\linewidth]{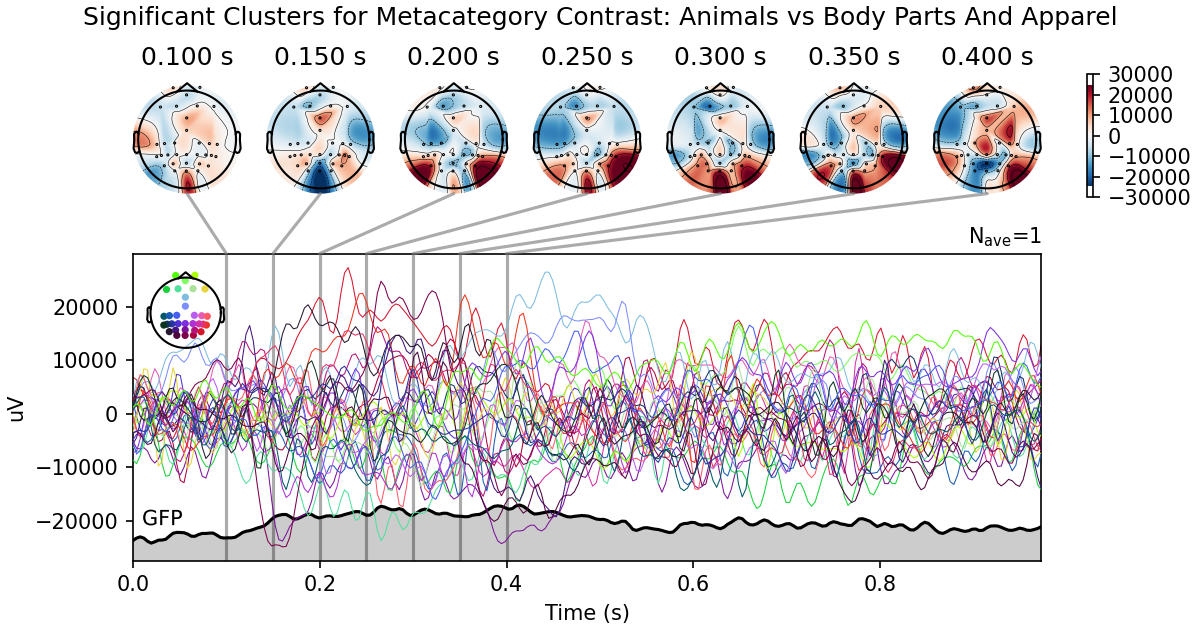}
    \caption{Cluster analyses results for the contrast between Animals and Body Parts/Apparel. Yellow shaded area show times where difference between signals was significant. White dots in topographical maps represent significant electrodes which were significantly different for those time periods. Gray shaded area represents change in Global Field Power (GFP) over time. }
    \label{fig:cluster3}
\end{figure}

\begin{figure}[!htb]
    \centering
    \includegraphics[width=\linewidth]{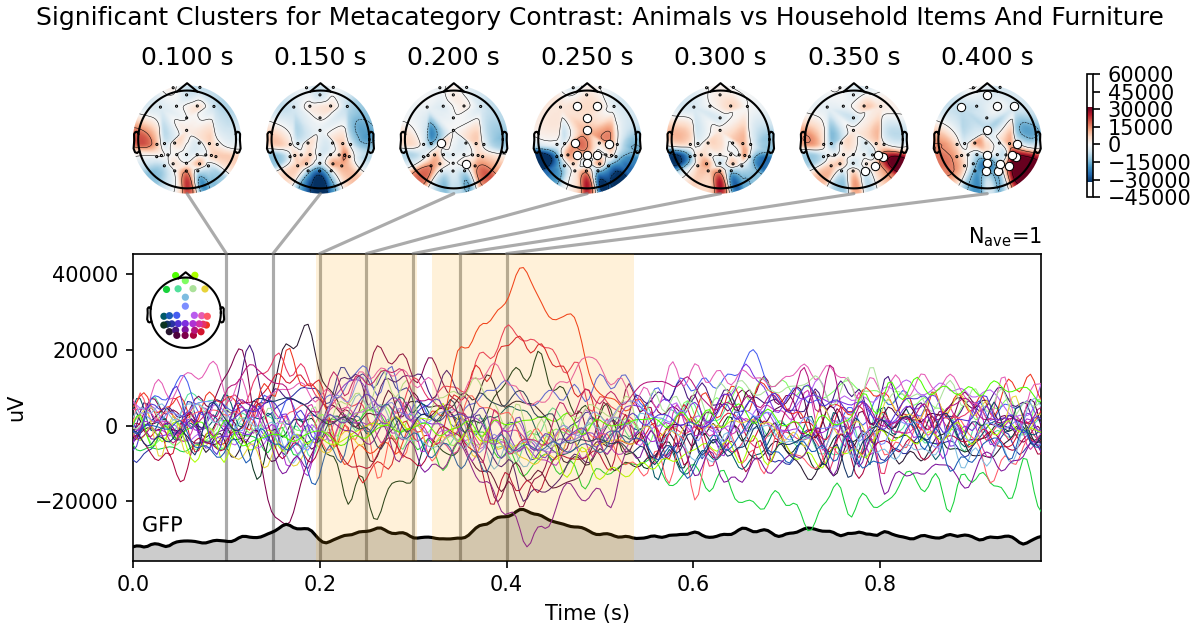}
    \caption{Cluster analyses results for the contrast between Animals and Household Items/Furniture. Yellow shaded area show times where difference between signals was significant. White dots in topographical maps represent significant electrodes which were significantly different for those time periods. Gray shaded area represents change in Global Field Power (GFP) over time. }
    \label{fig:cluster4}
\end{figure}

\begin{figure}[!htb]
    \centering
    \includegraphics[width=\linewidth]{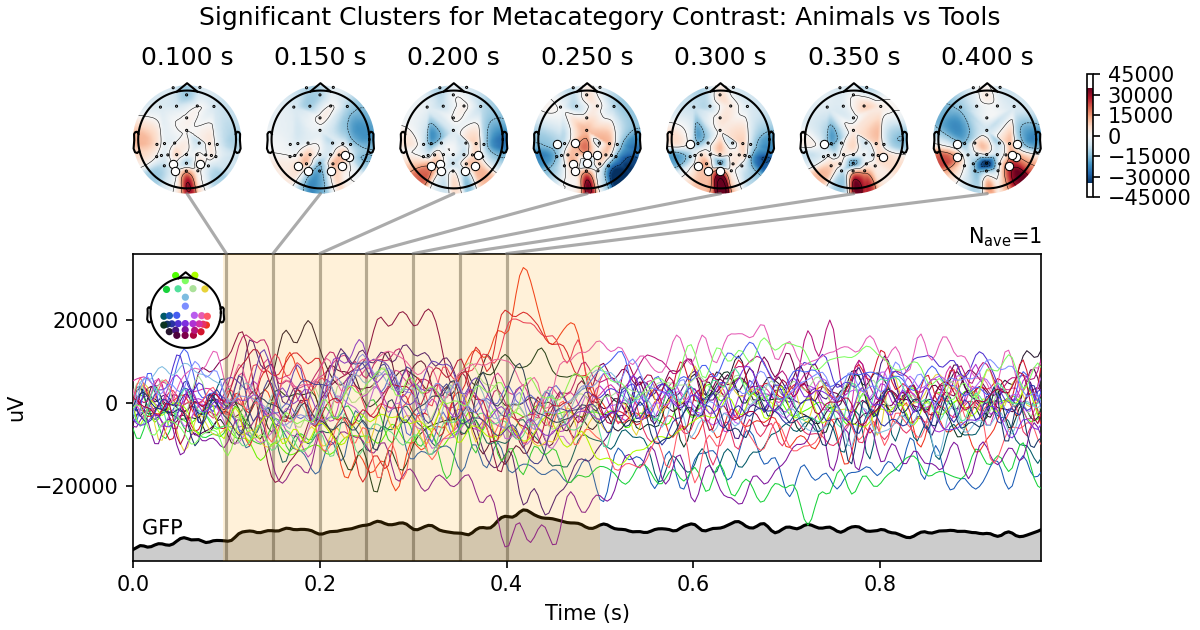}
    \caption{Cluster analyses results for the contrast between Animals and Tools. Yellow shaded area show times where difference between signals was significant. White dots in topographical maps represent significant electrodes which were significantly different for those time periods. Gray shaded area represents change in Global Field Power (GFP) over time. }
    \label{fig:cluster5}
\end{figure}

\begin{figure}[!htb]
    \centering
    \includegraphics[width=\linewidth]{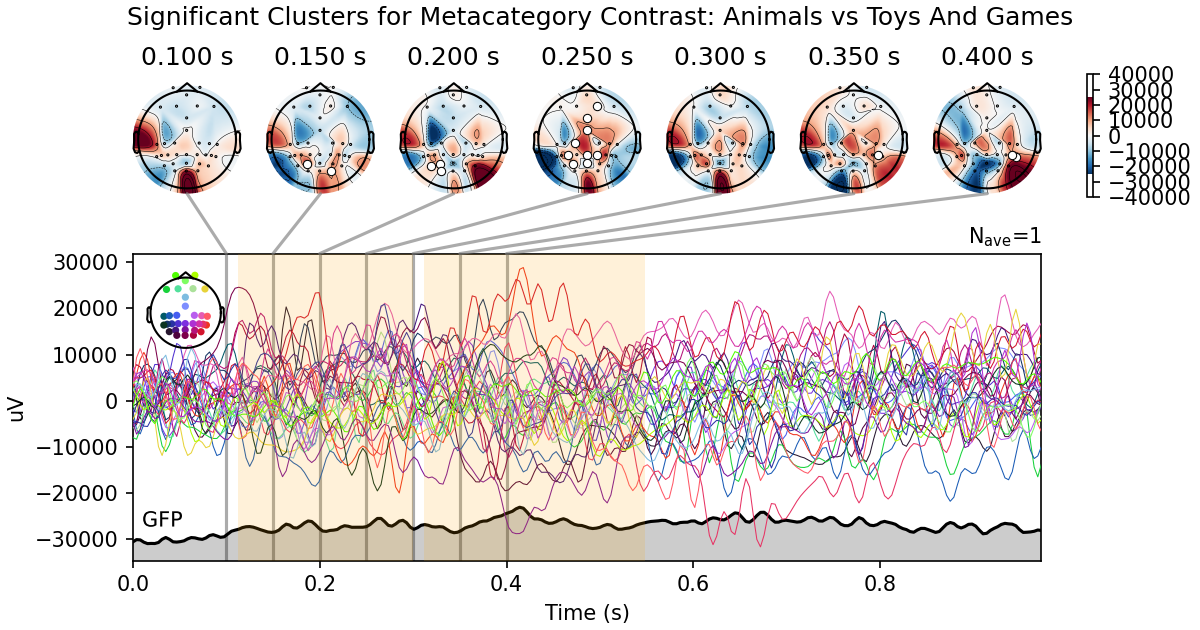}
    \caption{Cluster analyses results for the contrast between Animals and Toys/Games. Yellow shaded area show times where difference between signals was significant. White dots in topographical maps represent significant electrodes which were significantly different for those time periods. Gray shaded area represents change in Global Field Power (GFP) over time. }
    \label{fig:cluster6}
\end{figure}

\begin{figure}[!htb]
    \centering
    \includegraphics[width=\linewidth]{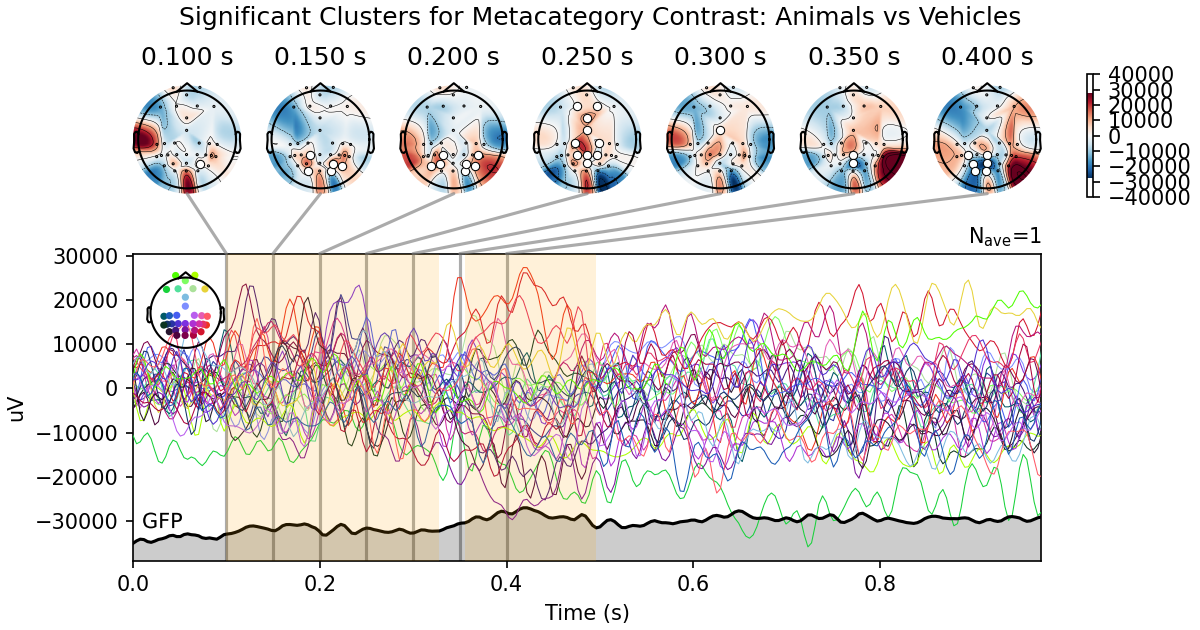}
    \caption{Cluster analyses results for the contrast between Animals and Vehicles. Yellow shaded area show times where difference between signals was significant. White dots in topographical maps represent significant electrodes which were significantly different for those time periods. Gray shaded area represents change in Global Field Power (GFP) over time. }
    \label{fig:cluster7}
\end{figure}

\begin{figure}[!htb]
    \centering
    \includegraphics[width=\linewidth]{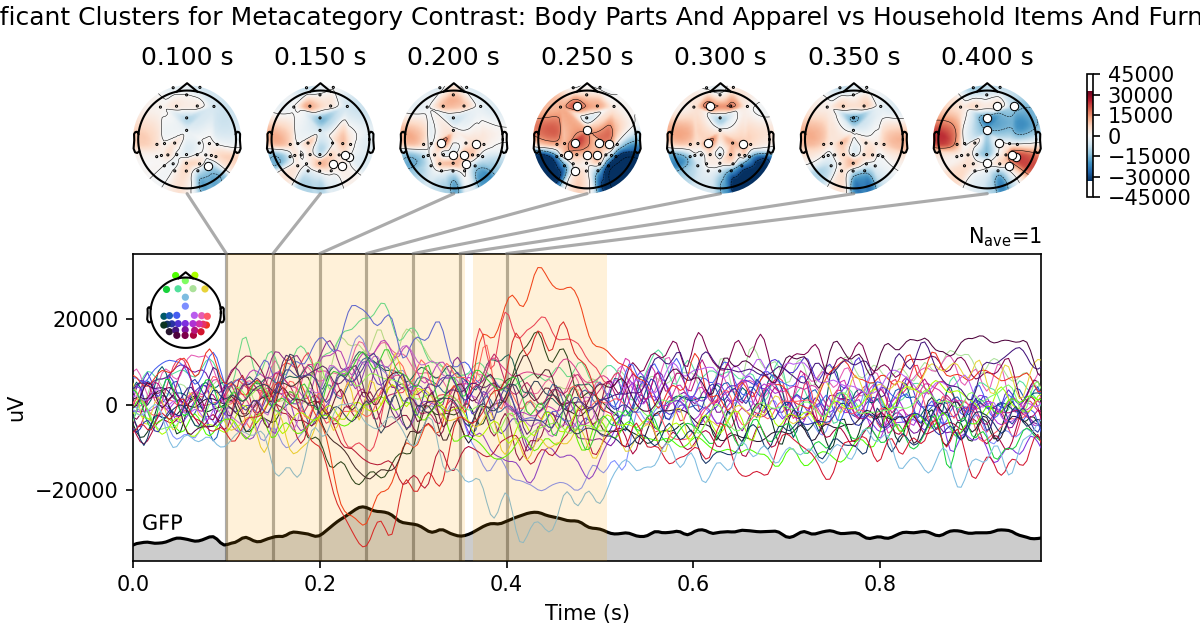}
    \caption{Cluster analyses results for the contrast between Body Parts/Apparel and Household Items/Furniture. Yellow shaded area show times where difference between signals was significant. White dots in topographical maps represent significant electrodes which were significantly different for those time periods. Gray shaded area represents change in Global Field Power (GFP) over time. }
    \label{fig:cluster8}
\end{figure}

\begin{figure}[!htb]
    \centering
    \includegraphics[width=\linewidth]{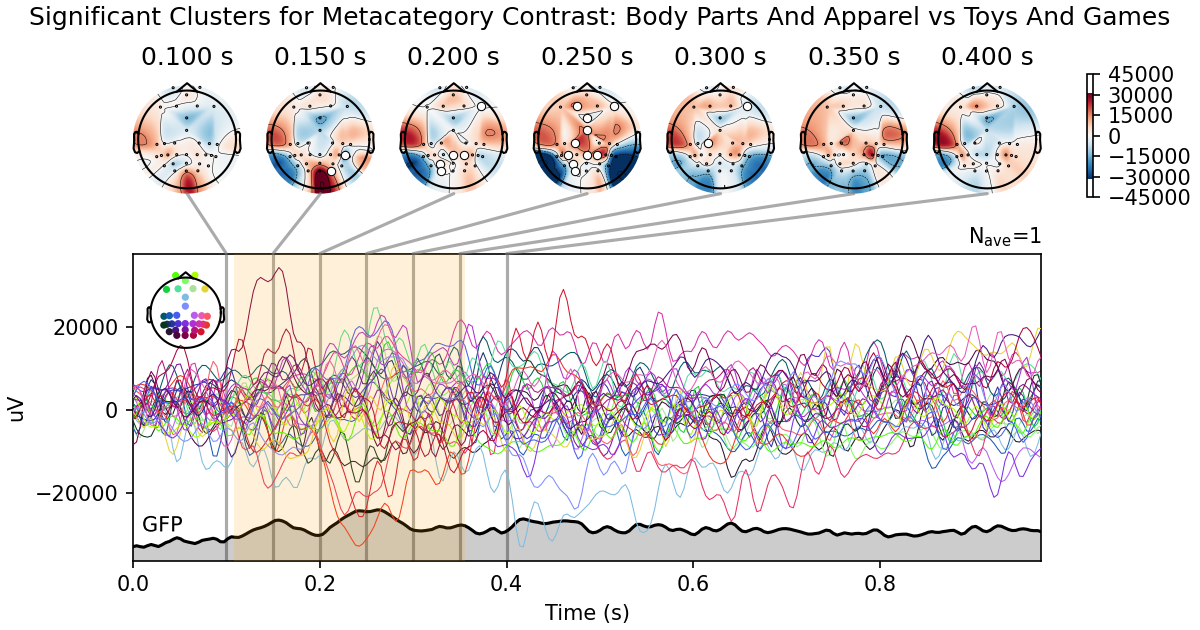}
    \caption{Cluster analyses results for the contrast between Body Parts/Apparel and Toys/Games. Yellow shaded area show times where difference between signals was significant. White dots in topographical maps represent significant electrodes which were significantly different for those time periods. Gray shaded area represents change in Global Field Power (GFP) over time. }
    \label{fig:cluster9}
\end{figure}

\begin{figure}[!htb]
    \centering
    \includegraphics[width=\linewidth]{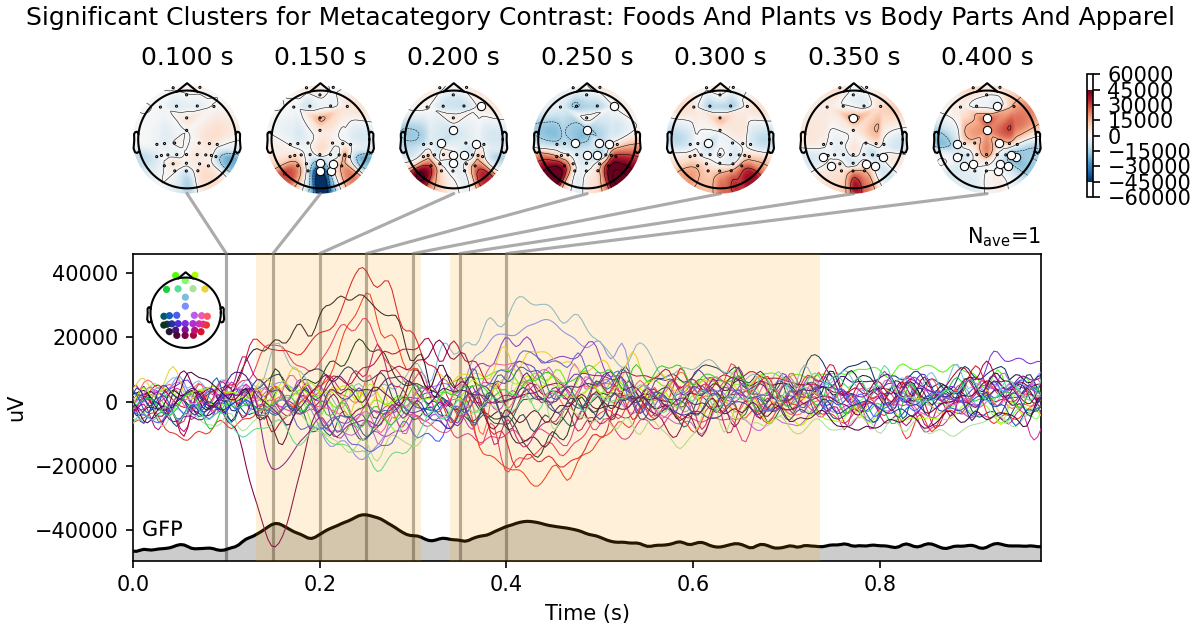}
    \caption{Cluster analyses results Foods/Plants and Body Parts/Apparel. Yellow shaded area show times where difference between signals was significant. White dots in topographical maps represent significant electrodes which were significantly different for those time periods. Gray shaded area represents change in Global Field Power (GFP) over time. }
    \label{fig:cluster10}
\end{figure}

\begin{figure}[!htb]
    \centering
    \includegraphics[width=\linewidth]{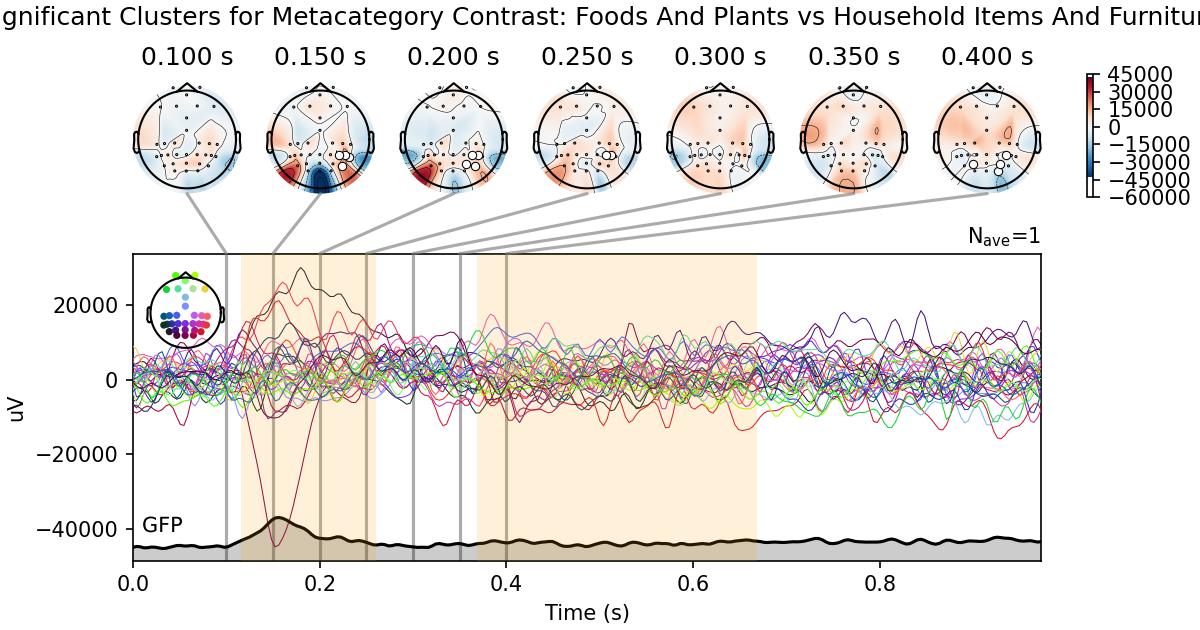}
    \caption{Cluster analyses results for the contrast between Foods/Plants and Household Items/Furniture. Yellow shaded area show times where difference between signals was significant. White dots in topographical maps represent significant electrodes which were significantly different for those time periods. Gray shaded area represents change in Global Field Power (GFP) over time. }
    \label{fig:cluster11}
\end{figure}

\begin{figure}[!htb]
    \centering
    \includegraphics[width=\linewidth]{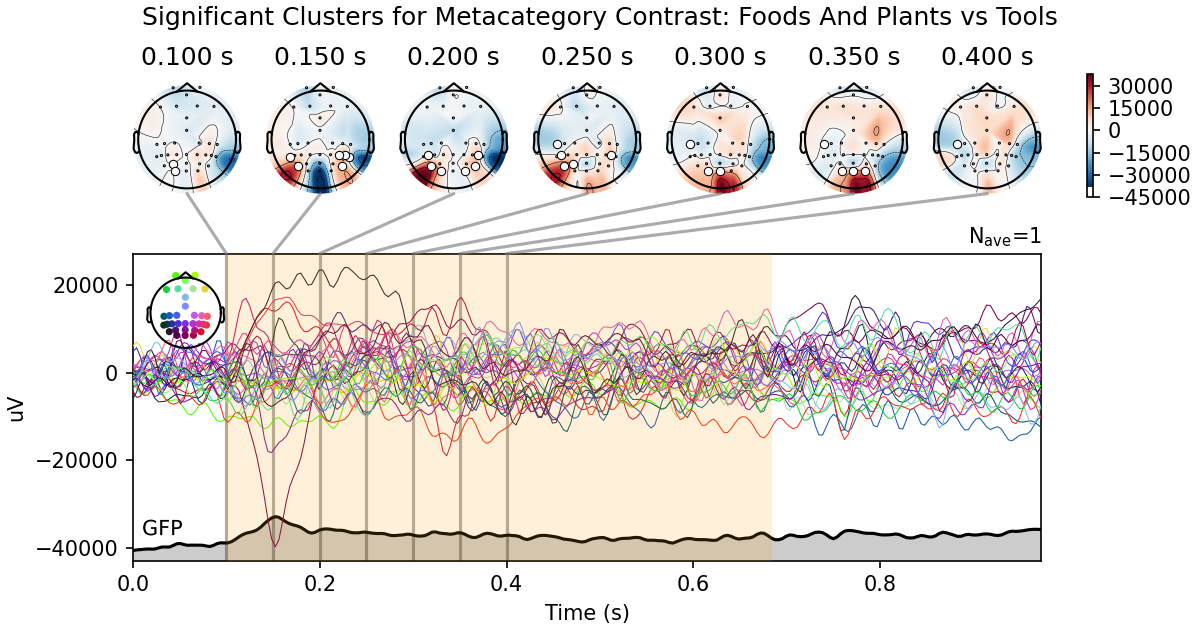}
    \caption{Cluster analyses results for the contrast between Foods/Plants and Tools. Yellow shaded area show times where difference between signals was significant. White dots in topographical maps represent significant electrodes which were significantly different for those time periods. Gray shaded area represents change in Global Field Power (GFP) over time. }
    \label{fig:cluster12}
\end{figure}

\begin{figure}[!htb]
    \centering
    \includegraphics[width=\linewidth]{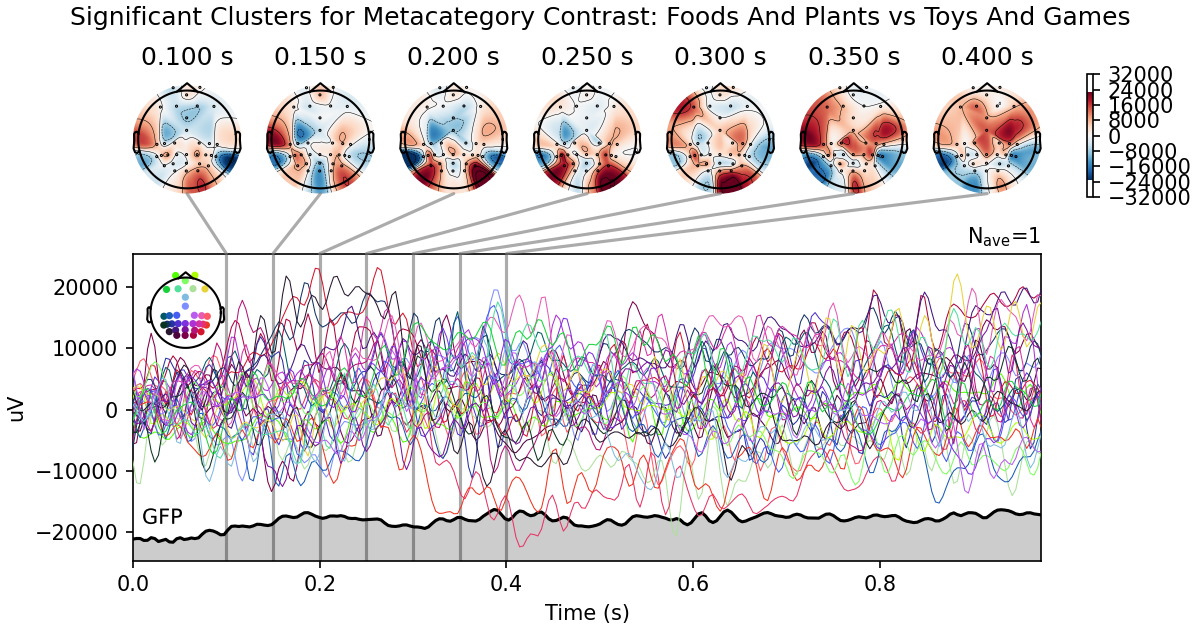}
    \caption{Cluster analyses results for the contrast between Foods/Plants and Toys/Games. Yellow shaded area show times where difference between signals was significant. White dots in topographical maps represent significant electrodes which were significantly different for those time periods. Gray shaded area represents change in Global Field Power (GFP) over time. }
    \label{fig:cluster13}
\end{figure}

\begin{figure}[!htb]
    \centering
    \includegraphics[width=\linewidth]{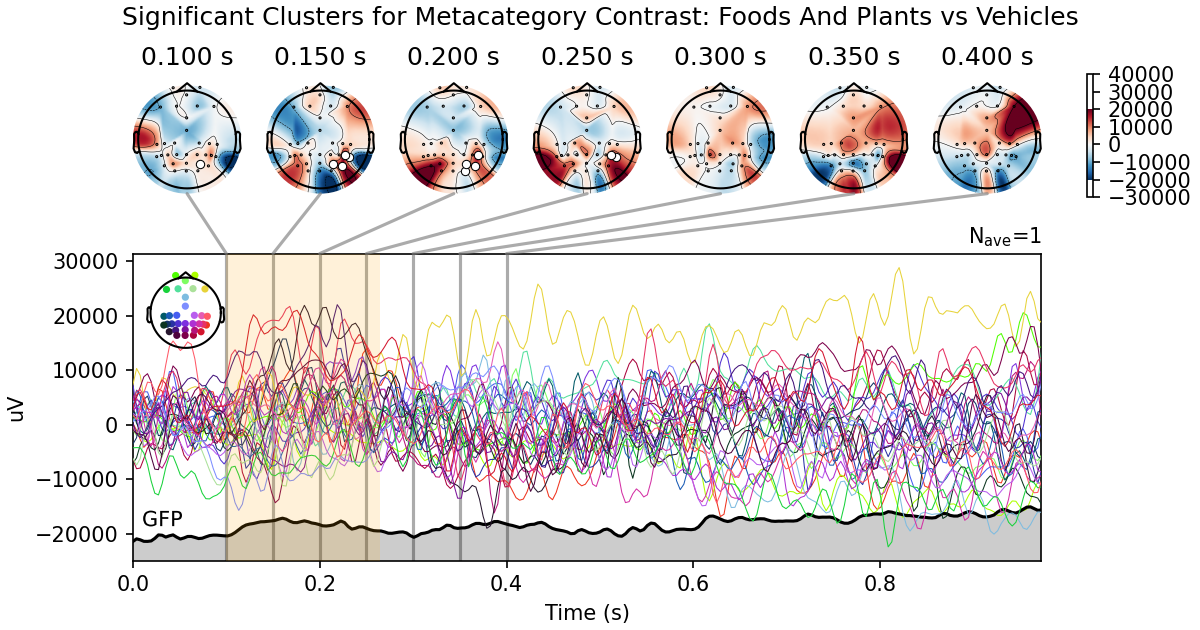}
    \caption{Cluster analyses results  for the contrast between Foods/Plants and Vehicles. Yellow shaded area show times where difference between signals was significant. White dots in topographical maps represent significant electrodes which were significantly different for those time periods. Gray shaded area represents change in Global Field Power (GFP) over time. }
    \label{fig:cluster14}
\end{figure}

\begin{figure}[!htb]
    \centering
    \includegraphics[width=\linewidth]{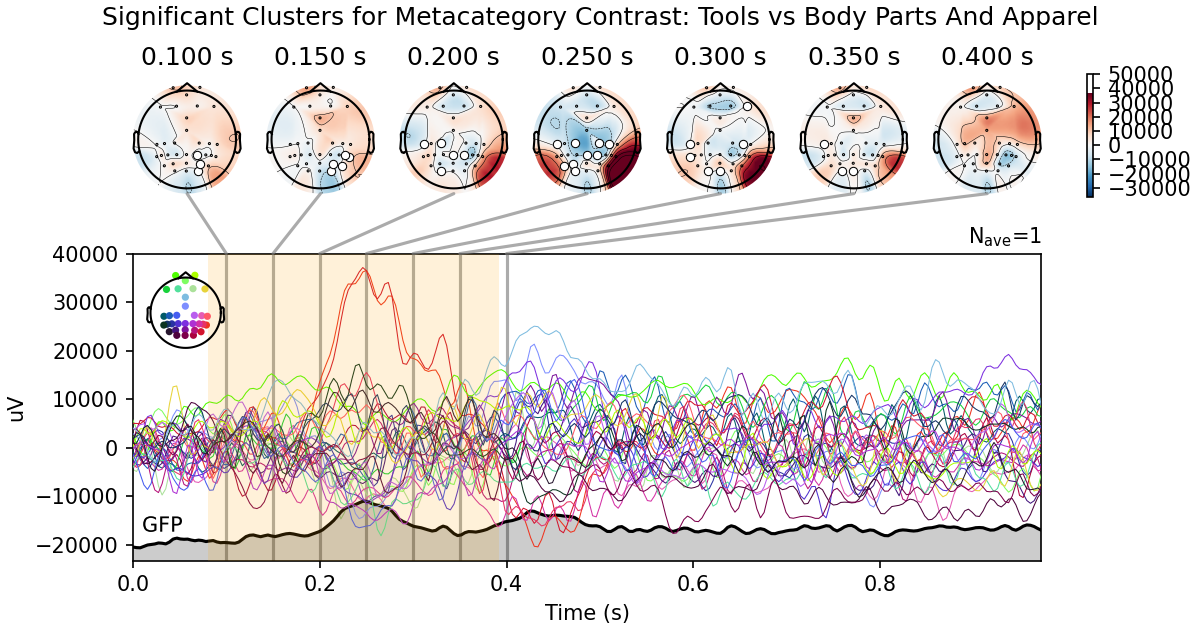}
    \caption{Cluster analyses results  for the contrast between Tools and Body Parts/Apparel. Yellow shaded area show times where difference between signals was significant. White dots in topographical maps represent significant electrodes which were significantly different for those time periods. Gray shaded area represents change in Global Field Power (GFP) over time. }
    \label{fig:cluster15}
\end{figure}

\begin{figure}[!htb]
    \centering
    \includegraphics[width=\linewidth]{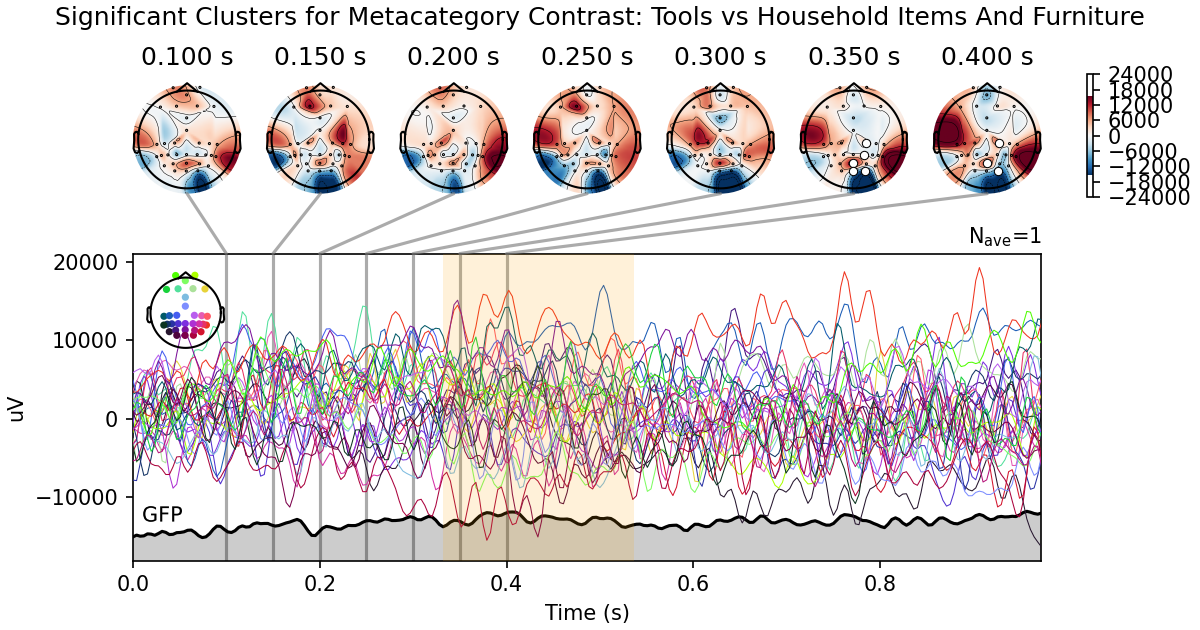}
    \caption{Cluster analyses results  for the contrast between Tools and Household Items/Furniture. Yellow shaded area show times where difference between signals was significant. White dots in topographical maps represent significant electrodes which were significantly different for those time periods. Gray shaded area represents change in Global Field Power (GFP) over time. }
    \label{fig:cluster16}
\end{figure}

\begin{figure}[!htb]
    \centering
    \includegraphics[width=\linewidth]{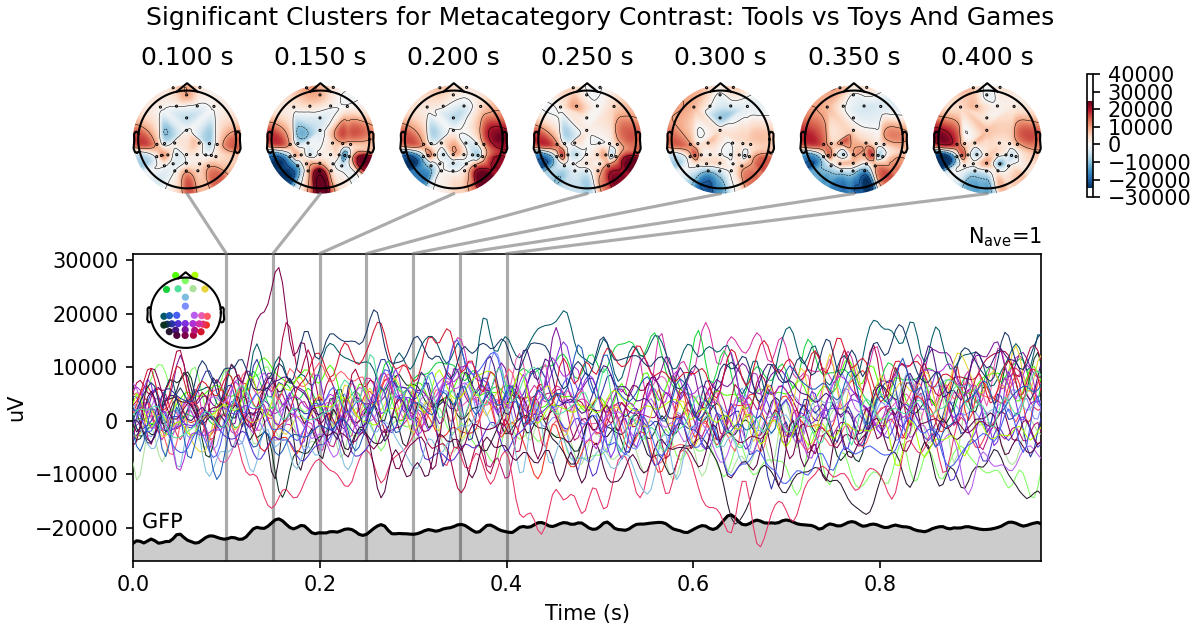}
    \caption{Cluster analyses results for the contrast between Tools and Toys/Games. Yellow shaded area show times where difference between signals was significant. White dots in topographical maps represent significant electrodes which were significantly different for those time periods. Gray shaded area represents change in Global Field Power (GFP) over time. }
    \label{fig:cluster17}
\end{figure}
\begin{figure}[!htb]
    \centering
    \includegraphics[width=\linewidth]{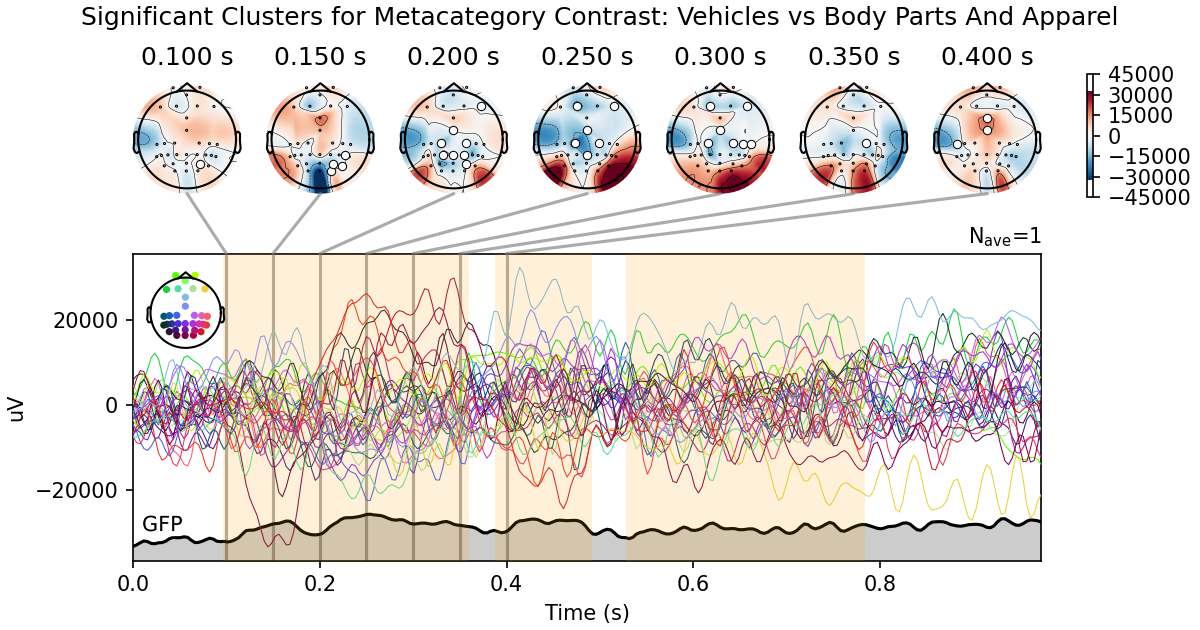}
    \caption{Cluster analyses results for the contrast between Vehicles and Body Parts/Apparel. Yellow shaded area show times where difference between signals was significant. White dots in topographical maps represent significant electrodes which were significantly different for those time periods. Gray shaded area represents change in Global Field Power (GFP) over time. }
    \label{fig:cluster18}
\end{figure}
\begin{figure}[!htb]
    \centering
    \includegraphics[width=\linewidth]{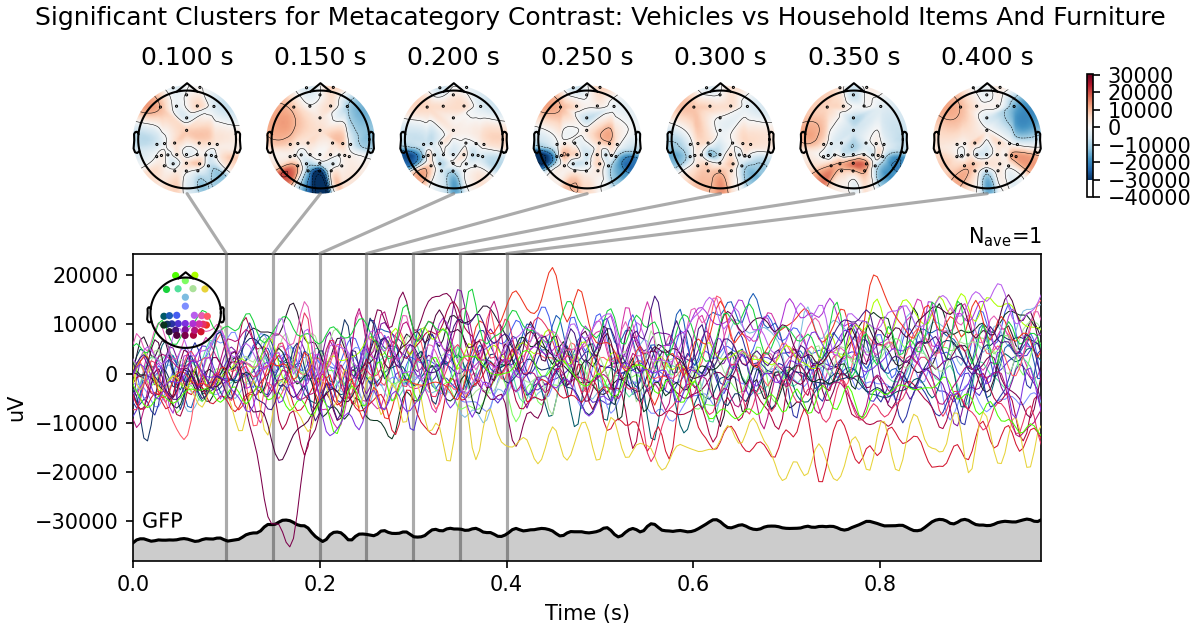}
    \caption{Cluster analyses results for the contrast between Vehicles and Household Items/Furniture. Yellow shaded area show times where difference between signals was significant. White dots in topographical maps represent significant electrodes which were significantly different for those time periods. Gray shaded area represents change in Global Field Power (GFP) over time. }
    \label{fig:cluster19}
\end{figure}
\begin{figure}[!htb]
    \centering
    \includegraphics[width=\linewidth]{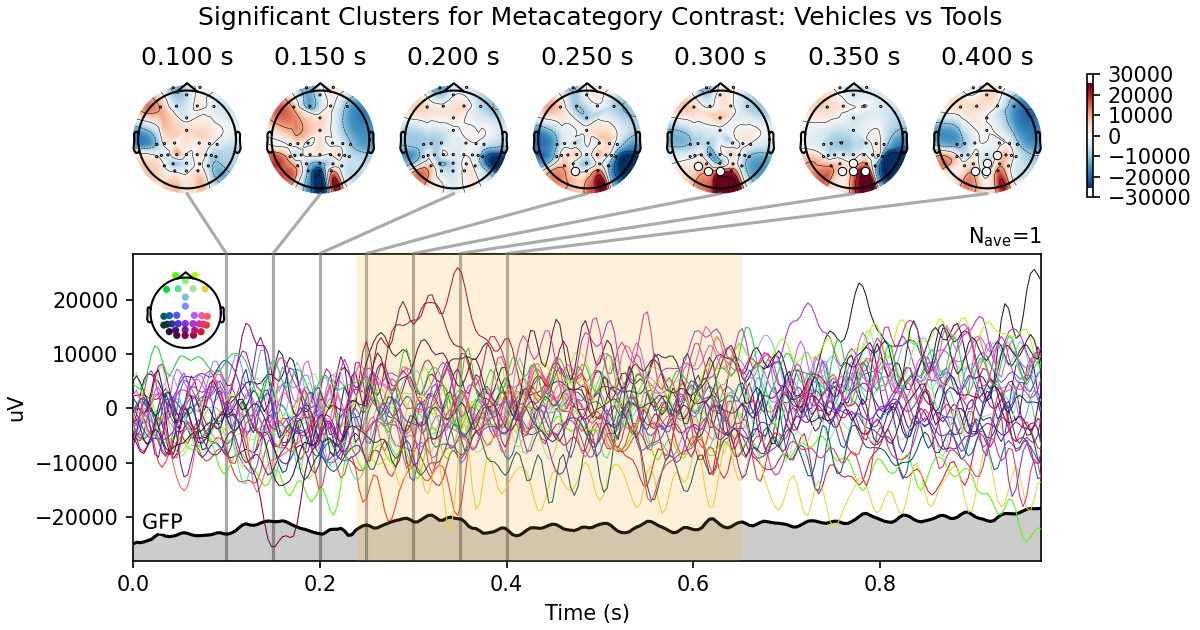}
    \caption{Cluster analyses results for the contrast between Vehicles and Tools. Yellow shaded area show times where difference between signals was significant. White dots in topographical maps represent significant electrodes which were significantly different for those time periods. Gray shaded area represents change in Global Field Power (GFP) over time. }
    \label{fig:cluster20}
\end{figure}
\begin{figure}[!htb]
    \centering
    \includegraphics[width=\linewidth]{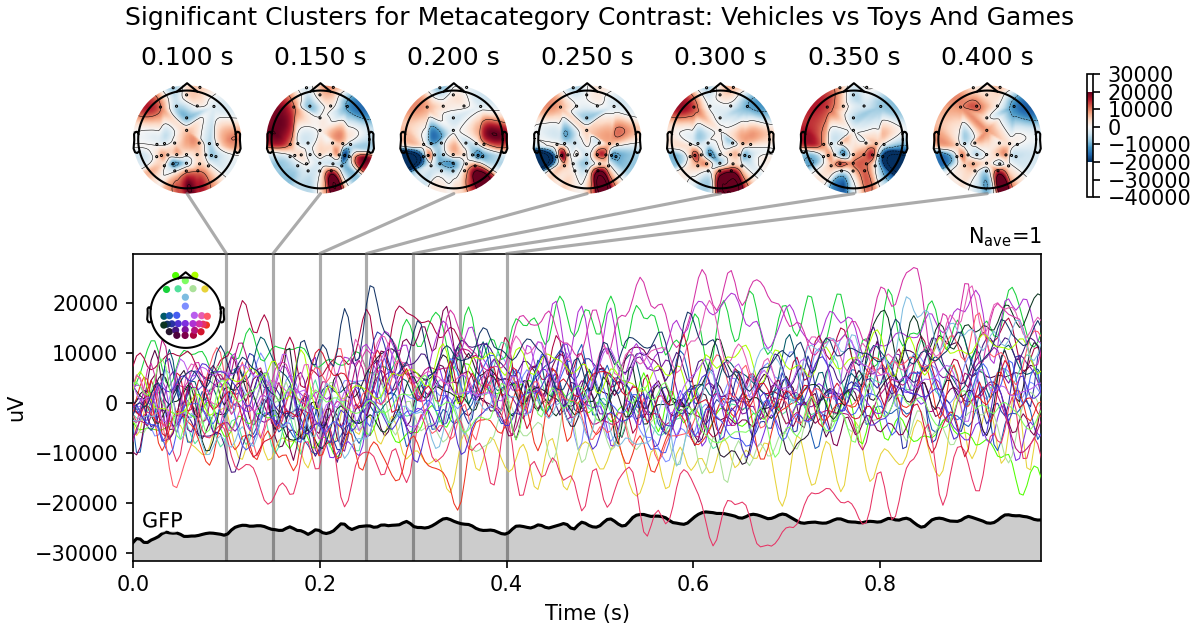}
    \caption{Cluster analyses results for the contrast between Vehicles and Toys/Games. Yellow shaded area show times where difference between signals was significant. White dots in topographical maps represent significant electrodes which were significantly different for those time periods. Gray shaded area represents change in Global Field Power (GFP) over time. }
    \label{fig:cluster21}
\end{figure}
\begin{figure}[!htb]
    \centering
    \includegraphics[width=\linewidth]{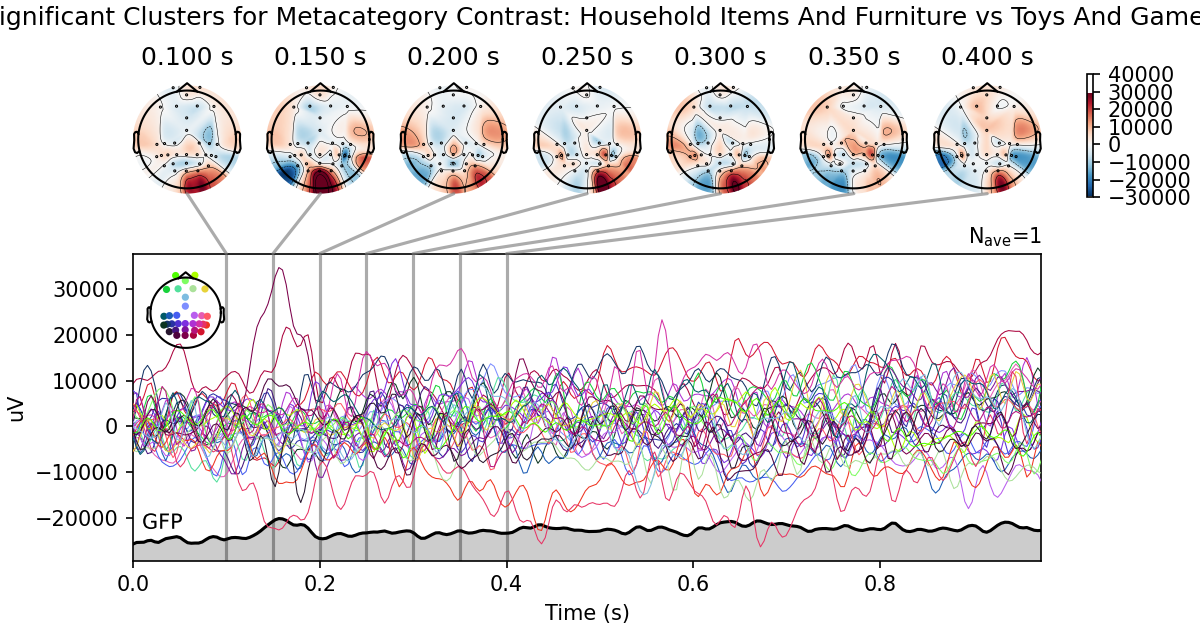}
    \caption{Cluster analyses results for the contrast between Household Items/Furniture and Toys/Games. Yellow shaded area show times where difference between signals was significant. White dots in topographical maps represent significant electrodes which were significantly different for those time periods. Gray shaded area represents change in Global Field Power (GFP) over time. }
    \label{fig:cluster22}
\end{figure}

\FloatBarrier 

\subsection{Saliency Maps and Activation Maximization}
\label{app:saliency}
\begin{figure}[!htb]
\centering
\includegraphics[width=0.8\textwidth]{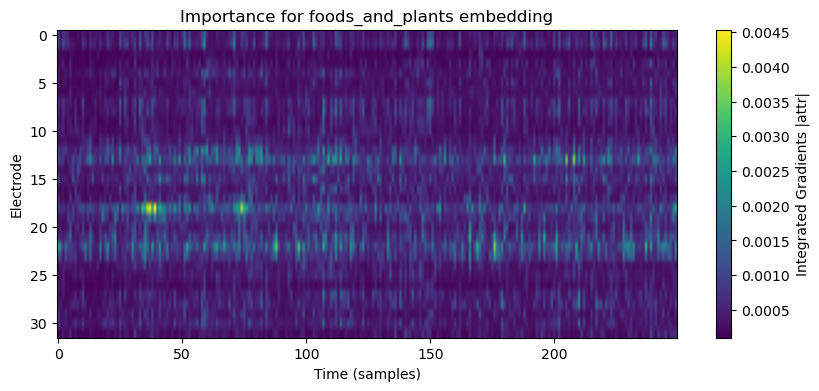}
\vspace{-5pt}
\caption{Raw Integrated-Gradients saliency matrix for the foods and plants embedding. Color indicates the absolute attribution assigned to each electrode-by-time sample pair; a few sparse, high-intensity pixels mark the model’s most influential spatiotemporal features.}
\label{fig:saliency_map}
\end{figure}

% --------------------------------------------------------------------------
% Saliency-map computation
% --------------------------------------------------------------------------
To compute the saliency maps shown in Figure \ref{fig:saliency_map} we applied
Integrated Gradients \citep{sundararajan2017axiomatic} with Captum 0.7.0 to a
frozen EEG$\!\to\!$CLIP encoder (ENIGMA).
For each meta-category (animals, household items / furniture, foods / plants,
tools) we first averaged the CLIP-ViT-L/14 embeddings of all training images
in that category; the resulting 1024-dimensional vector
$\mathbf{c}$ served as the optimization target.  
During attribution we maximized the cosine similarity between the model’s
predicted embedding $f_{\theta}(\mathbf{X},s)$ for an EEG trial
$\mathbf{X}\!\in\!\mathbb{R}^{C\times T}$ ($C{=}32$, $T{=}250$) from subject
$s$ and the category vector:

\begin{equation}
g\bigl(\mathbf{X}\bigr)
\;=\;
\cos\!\bigl(f_{\theta}(\mathbf{X},s),\,\mathbf{c}\bigr)
\;=\;
\frac{f_{\theta}(\mathbf{X},s)^{\top}\mathbf{c}}
     {\lVert f_{\theta}(\mathbf{X},s)\rVert_{2}\,
      \lVert\mathbf{c}\rVert_{2}},
\label{eq:forward_hook}
\end{equation}

using the subject-averaged grand-average trace
$\bar{\mathbf{X}}$ as the IG baseline.  
We ran 64 integration steps with an internal batch size of 32 and took the
unsigned attributions.

\begin{figure}[!htb]
\centering
\includegraphics[width=0.8\textwidth]{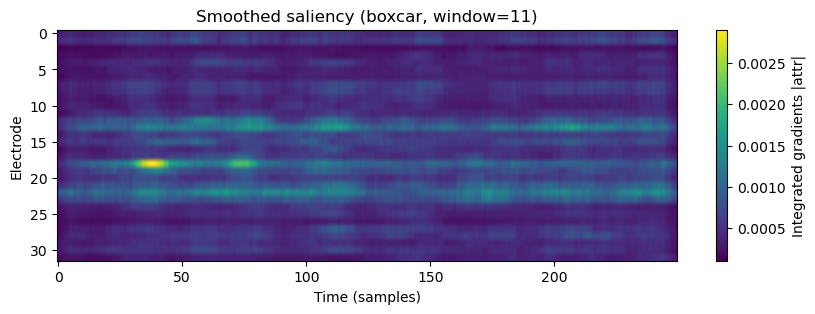}
\vspace{-5pt}
\caption{The same saliency matrix after convolution with an 11-sample boxcar kernel. Temporal smoothing suppresses noise and reveals a coherent band of elevated importance over mid-posterior channels between $\approx160$ ms and 300 ms post-stimulus.}
\label{fig:saliency_smooth}
\end{figure}

The per-trial saliencies were then accumulated over all $N$ training trials to
form a single importance tensor

\begin{equation}
\mathbf{S}
\;=\;
\frac{1}{N}\sum_{b=1}^{N}
\Bigl\lvert
\mathrm{IG}\!\bigl(\mathbf{X}^{(b)};\,g,\,\bar{\mathbf{X}}\bigr)
\Bigr\rvert,
\label{eq:saliency_accum}
\end{equation}

yielding $\mathbf{S}\!\in\!\mathbb{R}^{C\times T}$, the saliency map plotted in Figure \ref{fig:saliency_map}. We convolved each channel with an 11-sample (\,$\pm22$ ms) boxcar kernel to suppress high-frequency noise, converted the smoothed tensor into an \texttt{mne.EvokedArray} (Figure \ref{fig:saliency_smooth}), and plotted topographical maps at representative latencies on the standard 10–20 montage (Figure \ref{fig:saliency}).   Across categories the maps show a consistent attribution peak over occipital sensors between 160 ms and 300 ms post-stimulus, implicating early visual cortical activity.

We also visualized activation maximization for the ENIGMA architecture, by optimizing the input signal itself to synthesize an EEG pattern that most strongly activates a chosen semantic direction (e.g., “animals”). A learnable tensor initialized to zeros (shape = 1 × 32 × 250) was re-parameterized through a hyperbolic-tangent transform to keep values in (-1, 1). We maximized the cosine similarity between the model output and the target category vector while imposing an L1 sparsity penalty ($\lambda = 10^3$) to obtain interpretable traces. Optimization used Adam with a learning rate of $10^2$ for 2000 iterations on a single NVIDIA 3090 GPU.

\begin{figure}[!htb]
\centering
\includegraphics[width=0.8\textwidth]{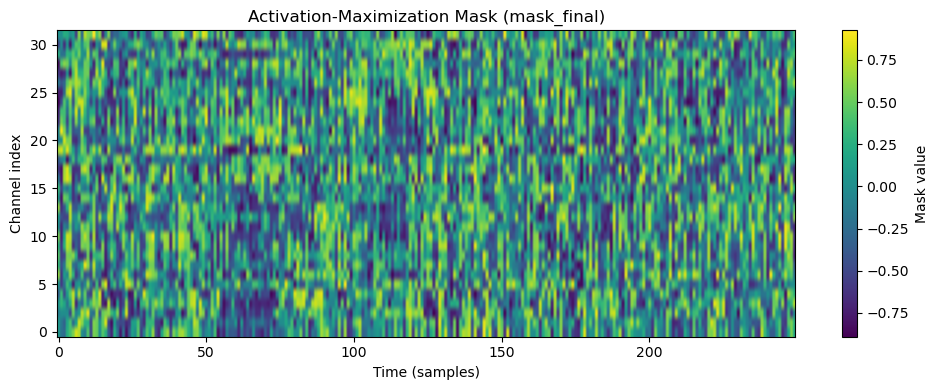}
\vspace{-5pt}
\caption{Activation-maximization mask produced by directly optimizing the input EEG to maximize cosine similarity with the foods and plants semantic direction. Yellow/green values enhance, and purple values suppress, the model’s activation, together outlining the composite pattern that most strongly drives the decoder.}
\label{fig:activation_maximisation}
\end{figure}

% \FloatBarrier
% \clearpage 

% \section{Concurrent Submission}

% \includepdf[pages=-]{figures/ENIGMA_Submission.pdf}

%% file: main.bbl
\begin{thebibliography}{10}

\bibitem{allen_massive_2022}
Emily~J. Allen, Ghislain St-Yves, Yihan Wu, Jesse~L. Breedlove, Jacob~S. Prince, Logan~T. Dowdle, Matthias Nau, Brad Caron, Franco Pestilli, Ian Charest, J.~Benjamin Hutchinson, Thomas Naselaris, and Kendrick Kay.
\newblock A massive {7T} {fMRI} dataset to bridge cognitive neuroscience and artificial intelligence.
\newblock {\em Nature Neuroscience}, 25(1):116--126, January 2022.

\bibitem{ENIGMA}
Anonymous.
\newblock Enigma: A unified lightweight eeg-to-image model for multi-subject visual decoding.
\newblock In Review, see Appendix B., 2025.

\bibitem{badcock2013validation}
Nicholas~A Badcock, Petroula Mousikou, Yatin Mahajan, Peter De~Lissa, Johnson Thie, and Genevieve McArthur.
\newblock Validation of the emotiv epoc{\textregistered} eeg gaming system for measuring research quality auditory erps.
\newblock {\em PeerJ}, 1:e38, 2013.

\bibitem{badcock2015validation}
Nicholas~A Badcock, Kathryn~A Preece, Bianca de~Wit, Katharine Glenn, Nora Fieder, Johnson Thie, and Genevieve McArthur.
\newblock Validation of the emotiv epoc eeg system for research quality auditory event-related potentials in children.
\newblock {\em PeerJ}, 3:e907, 2015.

\bibitem{banville2025scaling}
Hubert Banville, Yohann Benchetrit, St{\'e}phane d'Ascoli, J{\'e}r{\'e}my Rapin, and Jean-R{\'e}mi King.
\newblock Scaling laws for decoding images from brain activity.
\newblock {\em arXiv preprint arXiv:2501.15322}, 2025.

\bibitem{Banville2021}
Hubert Banville, Yohann Benchetrit, Stéphane d'Ascoli, Jérémy Rapin, and Jean-Rémi King.
\newblock Uncovering the structure of clinical eeg signals with self-supervised learning.
\newblock {\em Journal of Neural Engineering}, 18(4):046020, 2021.

\bibitem{bentin1996electrophysiological}
Shlomo Bentin, Truett Allison, Aina Puce, Erik Perez, and Gregory McCarthy.
\newblock Electrophysiological studies of face perception in humans.
\newblock {\em Journal of cognitive neuroscience}, 8(6):551--565, 1996.

\bibitem{bhavnani2022acceptability}
Supriya Bhavnani, Dhanya Parameshwaran, Kamal~Kant Sharma, Debarati Mukherjee, Gauri Divan, Vikram Patel, and Tara~C Thiagarajan.
\newblock The acceptability, feasibility, and utility of portable electroencephalography to study resting-state neurophysiology in rural communities.
\newblock {\em Frontiers in human neuroscience}, 16:802764, 2022.

\bibitem{brown2020language}
Tom Brown, Benjamin Mann, Nick Ryder, Melanie Subbiah, Jared~D Kaplan, Prafulla Dhariwal, Arvind Neelakantan, Pranav Shyam, Girish Sastry, Amanda Askell, et~al.
\newblock Language models are few-shot learners.
\newblock {\em Advances in neural information processing systems}, 33:1877--1901, 2020.

\bibitem{caron_unsupervised_2021}
Mathilde Caron, Ishan Misra, Julien Mairal, Priya Goyal, Piotr Bojanowski, and Armand Joulin.
\newblock Unsupervised learning of visual features by contrasting cluster assignments.
\newblock {\em CoRR}, abs/2006.09882, 2020.

\bibitem{chang_bold5000_2019}
Nadine Chang, John~A. Pyles, Austin Marcus, Abhinav Gupta, Michael~J. Tarr, and Elissa~M. Aminoff.
\newblock {BOLD5000}, a public {fMRI} dataset while viewing 5000 visual images.
\newblock {\em Scientific Data}, 6(1):49, May 2019.
\newblock Number: 1 Publisher: Nature Publishing Group.

\bibitem{chen2023structure}
Zijiao Chen, Jonathan Xu, Jiaxin Qing, Ruilin Li, and Juan~Helen Zhou.
\newblock Structure-preserved image reconstruction from brain recordings, 2023.

\bibitem{Cichy2014}
Radoslaw~M Cichy, Dimitrios Pantazis, and Aude Oliva.
\newblock Resolving human object recognition in space and time.
\newblock {\em Nature Neuroscience}, 17(3):455--462, 2014.

\bibitem{cichy2014resolving}
Radoslaw~Martin Cichy, Dimitrios Pantazis, and Aude Oliva.
\newblock Resolving human object recognition in space and time.
\newblock {\em Nature neuroscience}, 17(3):455--462, 2014.

\bibitem{deng2009imagenet}
Jia Deng, Wei Dong, Richard Socher, Li-Jia Li, Kai Li, and Li~Fei-Fei.
\newblock Imagenet: A large-scale hierarchical image database.
\newblock In {\em 2009 IEEE conference on computer vision and pattern recognition}, pages 248--255. Ieee, 2009.

\bibitem{devlin2019bert}
Jacob Devlin, Ming-Wei Chang, Kenton Lee, and Kristina Toutanova.
\newblock Bert: Pre-training of deep bidirectional transformers for language understanding.
\newblock In {\em Proceedings of the 2019 conference of the North American chapter of the association for computational linguistics: human language technologies, volume 1 (long and short papers)}, pages 4171--4186, 2019.

\bibitem{duvinage2012p300}
Matthieu Duvinage, Thierry Castermans, Thierry Dutoit, Mathieu Petieau, Thomas Hoellinger, Caty~De Saedeleer, K~Seetharaman, and G~Cheron.
\newblock A p300-based quantitative comparison between the emotiv epoc headset and a medical eeg device.
\newblock {\em Biomedical Engineering}, 765(1):2012--2764, 2012.

\bibitem{eimer2000face}
Martin Eimer.
\newblock The face-specific n170 component reflects late stages in the structural encoding of faces.
\newblock {\em Neuroreport}, 11(10):2319--2324, 2000.

\bibitem{Fei2024-perceptogram}
Teng Fei, Abhinav Uppal, Ian Jackson, Srinivas Ravishankar, David Wang, and Virginia~R. de~Sa.
\newblock {Perceptogram: Reconstructing Visual Percepts from EEG}.
\newblock {\em arXiv preprint arXiv:2404.01250}, 2024.
\newblock (extended version with additional analyses).

\bibitem{fernandez2022effect}
{\'A}lvaro Fern{\'a}ndez-Rodr{\'\i}guez, Aube Darves-Bornoz, Francisco Velasco-{\'A}lvarez, and Ricardo Ron-Angevin.
\newblock Effect of stimulus size in a visual erp-based bci under rsvp.
\newblock {\em Sensors}, 22(23):9505, 2022.

\bibitem{Gifford2022}
Alessandro~T. Gifford, Kshitij Dwivedi, Gemma Roig, and Radoslaw~M. Cichy.
\newblock A large and rich eeg dataset for modeling human visual object recognition.
\newblock {\em NeuroImage}, 264:119754, 2022.

\bibitem{sethics}
Emma~C. Gordon and Anil~K. Seth.
\newblock Ethical considerations for the use of brain–computer interfaces for cognitive enhancement.
\newblock {\em PLOS Biology}, 22(10):1--15, 10 2024.

\bibitem{Gramfort2013}
Alexandre Gramfort, Martin Luessi, Eric Larson, Denis~A. Engemann, Daniel Strohmeier, Christian Brodbeck, and et~al.
\newblock Meg and eeg data analysis with mne-python.
\newblock {\em Frontiers in Neuroscience}, 7:267, 2013.

\bibitem{Grootswagers2022}
Tijl Grootswagers, Ivy Zhou, Austin~K. Robinson, Michael~N. Hebart, and Thomas~A. Carlson.
\newblock Human eeg recordings for 1,854 concepts presented in rapid serial visual presentation streams.
\newblock {\em Scientific Data}, 9:3, 2022.

\bibitem{guggenmos2018multivariate}
Matthias Guggenmos, Philipp Sterzer, and Radoslaw~Martin Cichy.
\newblock Multivariate pattern analysis for meg: A comparison of dissimilarity measures.
\newblock {\em Neuroimage}, 173:434--447, 2018.

\bibitem{Hebart2019}
Michael~N. Hebart, Adam~H. Dickter, Alexis Kidder, Anna Corriveau, Cody Van~Wicklin, and Chris~I. Baker.
\newblock Things: A database of 1,854 object concepts and more than 26,000 naturalistic object images.
\newblock {\em PLOS ONE}, 14(10):e0223792, 2019.

\bibitem{ho2020denoising}
Jonathan Ho, Ajay Jain, and Pieter Abbeel.
\newblock Denoising diffusion probabilistic models.
\newblock {\em Advances in neural information processing systems}, 33:6840--6851, 2020.

\bibitem{holm2023contribution}
Eric~L{\"u}tzow Holm, Diego~Fern{\'a}ndez Slezak, and Enzo Tagliazucchi.
\newblock Contribution of image statistics and semantics in local vs. distributed eeg decoding of rapid serial visual presentation.
\newblock {\em bioRxiv}, pages 2023--09, 2023.

\bibitem{horikawa2017generic}
Tomoyasu Horikawa and Yukiyasu Kamitani.
\newblock Generic decoding of seen and imagined objects using hierarchical visual features.
\newblock {\em Nature communications}, 8(1):15037, 2017.

\bibitem{kneeland_brain-optimized_2023}
Reese Kneeland, Jordyn Ojeda, Ghislain St-Yves, and Thomas Naselaris.
\newblock Brain-optimized inference improves reconstructions of {fMRI} brain activity, December 2023.
\newblock arXiv:2312.07705 [cs, q-bio].

\bibitem{kneeland2023reconstructing}
Reese Kneeland, Jordyn Ojeda, Ghislain St-Yves, and Thomas Naselaris.
\newblock Reconstructing seen images from human brain activity via guided stochastic search.
\newblock In {\em Conference on Cognitive Computational Neuroscience}, 2023.

\bibitem{kneeland_second_2023}
Reese Kneeland, Jordyn Ojeda, Ghislain St-Yves, and Thomas Naselaris.
\newblock Second {Sight}: {Using} brain-optimized encoding models to align image distributions with human brain activity, June 2023.
\newblock arXiv:2306.00927 [cs, q-bio].

\bibitem{knierim2025advancing}
Michael~T Knierim, Christian Zimny, Gabriel Ivucic, and Tobias R{\"o}ddiger.
\newblock Advancing wearable bci: Headphone eeg for cognitive load detection in lab and field.
\newblock {\em Proceedings of the ACM on Interactive, Mobile, Wearable and Ubiquitous Technologies}, 9(1):1--26, 2025.

\bibitem{alexnet}
Alex Krizhevsky, Ilya Sutskever, and Geoffrey~E Hinton.
\newblock Imagenet classification with deep convolutional neural networks.
\newblock In F.~Pereira, C.J. Burges, L.~Bottou, and K.Q. Weinberger, editors, {\em Advances in Neural Information Processing Systems}, volume~25. Curran Associates, Inc., 2012.

\bibitem{Kutas1980}
Marta Kutas and Steven~A Hillyard.
\newblock Reading senseless sentences: Brain potentials reflect semantic incongruity.
\newblock {\em Science}, 207(4427):203--205, 1980.

\bibitem{Lawhern2018}
Vernon~J. Lawhern, Alex~J. Solon, Nicholas~R. Waytowich, Stacey~M. Gordon, Christine~P. Hung, and Brent~J. Lance.
\newblock Eegnet: a compact convolutional neural network for eeg-based brain--computer interfaces.
\newblock {\em Journal of Neural Engineering}, 15(5):056013, 2018.

\bibitem{Li2024}
Dongyang Li, Chen Wei, Shiying Li, Jiachen Zou, and Quanying Liu.
\newblock {Visual Decoding and Reconstruction via EEG Embeddings with Guided Diffusion}.
\newblock In {\em Advances in Neural Information Processing Systems (NeurIPS)}, 2024.

\bibitem{li2018training}
Ren Li, Jared~S Johansen, Hamad Ahmed, Thomas~V Ilyevsky, Ronnie~B Wilbur, Hari~M Bharadwaj, and Jeffrey~Mark Siskind.
\newblock Training on the test set? an analysis of spampinato et al.[31].
\newblock {\em arXiv preprint arXiv:1812.07697}, 2018.

\bibitem{li2020perils}
Ren Li, Jared~S Johansen, Hamad Ahmed, Thomas~V Ilyevsky, Ronnie~B Wilbur, Hari~M Bharadwaj, and Jeffrey~Mark Siskind.
\newblock The perils and pitfalls of block design for eeg classification experiments.
\newblock {\em IEEE Transactions on Pattern Analysis and Machine Intelligence}, 43(1):316--333, 2020.

\bibitem{luck2014introduction}
Steven~J Luck.
\newblock {\em An introduction to the event-related potential technique}.
\newblock MIT press, 2014.

\bibitem{maris2007nonparametric}
Eric Maris and Robert Oostenveld.
\newblock Nonparametric statistical testing of eeg-and meg-data.
\newblock {\em Journal of neuroscience methods}, 164(1):177--190, 2007.

\bibitem{michel2018eeg}
Christoph~M Michel and Thomas Koenig.
\newblock Eeg microstates as a tool for studying the temporal dynamics of whole-brain neuronal networks: a review.
\newblock {\em Neuroimage}, 180:577--593, 2018.

\bibitem{mikhaylov2024comparison}
Dmitry Mikhaylov, Muhammad Saeed, Mohamed Husain~Alhosani, and Yasser F.~Al~Wahedi.
\newblock Comparison of eeg signal spectral characteristics obtained with consumer-and research-grade devices.
\newblock {\em Sensors}, 24(24):8108, 2024.

\bibitem{niso2016omega}
Guiomar Niso, Christine Rogers, Jeremy~T Moreau, Li-Yuan Chen, Cecile Madjar, Samir Das, Elizabeth Bock, Fran{\c{c}}ois Tadel, Alan~C Evans, Pierre Jolicoeur, et~al.
\newblock Omega: the open meg archive.
\newblock {\em Neuroimage}, 124:1182--1187, 2016.

\bibitem{obeid2016temple}
Iyad Obeid and Joseph Picone.
\newblock The temple university hospital eeg data corpus.
\newblock {\em Frontiers in neuroscience}, 10:196, 2016.

\bibitem{ozcelik2023braindiffuser}
Furkan Ozcelik and Rufin VanRullen.
\newblock Natural scene reconstruction from {fMRI} signals using generative latent diffusion.
\newblock {\em Scientific Reports}, 13, 2023.

\bibitem{pfurtscheller1999event}
Gert Pfurtscheller and FH~Lopes Da~Silva.
\newblock Event-related eeg/meg synchronization and desynchronization: basic principles.
\newblock {\em Clinical neurophysiology}, 110(11):1842--1857, 1999.

\bibitem{Polich2007}
John Polich.
\newblock Updating p300: An integrative theory of p3a and p3b.
\newblock {\em Clinical Neurophysiology}, 118(10):2128--2148, 2007.

\bibitem{radford2021learning}
Alec Radford, Jong~Wook Kim, Chris Hallacy, Aditya Ramesh, Gabriel Goh, Sandhini Agarwal, Girish Sastry, Amanda Askell, Pamela Mishkin, Jack Clark, et~al.
\newblock Learning transferable visual models from natural language supervision.
\newblock In {\em International conference on machine learning}, pages 8748--8763. PmLR, 2021.

\bibitem{Roy2019}
Yannick Roy, Hubert Banville, Isabela Albuquerque, Alexandre Gramfort, Tiago~H. Falk, and Jocelyn Faubert.
\newblock Deep learning-based electroencephalography analysis: a systematic review.
\newblock {\em Journal of Neural Engineering}, 16(5):051001, 2019.

\bibitem{sabio2024scoping}
Joshua Sabio, Nikolas~S Williams, Genevieve~M McArthur, and Nicholas~A Badcock.
\newblock A scoping review on the use of consumer-grade eeg devices for research.
\newblock {\em Plos one}, 19(3):e0291186, 2024.

\bibitem{sato2024scaling}
Motoshige Sato, Kenichi Tomeoka, Ilya Horiguchi, Kai Arulkumaran, Ryota Kanai, and Shuntaro Sasai.
\newblock Scaling law in neural data: Non-invasive speech decoding with 175 hours of eeg data.
\newblock {\em arXiv preprint arXiv:2407.07595}, 2024.

\bibitem{Scotti2024MindEye2}
Paul~S. Scotti, Mihir Tripathy, Cesare Kadir~Torrico Villanueva, Reese Kneeland, Tong Chen, Ashutosh Narang, Charan Santhirasegaran, Jonathan Xu, Thomas Naselaris, Kenneth~A. Norman, and Tanishq~Mathew Abraham.
\newblock Mindeye2: shared-subject models enable fmri-to-image with 1 hour of data.
\newblock In {\em Proceedings of the 41st International Conference on Machine Learning}, 2024.

\bibitem{scotti_reconstructing_2023}
Paul~Steven Scotti, Atmadeep Banerjee, Jimmie Goode, Stepan Shabalin, Alex Nguyen, Cohen Ethan, Aidan~James Dempster, Nathalie Verlinde, Elad Yundler, David Weisberg, Kenneth Norman, and Tanishq~Mathew Abraham.
\newblock Reconstructing the mind's eye: f{MRI}-to-image with contrastive learning and diffusion priors.
\newblock In {\em Thirty-seventh Conference on Neural Information Processing Systems}, 2023.

\bibitem{Song2024}
Yonghao Song, Bingchuan Liu, Xiang Li, Nanlin Shi, Yijun Wang, and Xiaorong Gao.
\newblock {Decoding Natural Images from EEG for Object Recognition}.
\newblock In {\em Proceedings of the International Conference on Learning Representations (ICLR)}, 2024.

\bibitem{Spampinato2017}
Carlo Spampinato, Sebastiano Palazzo, Ignazio Kavasidis, Daniele Giordano, Nada Souly, and Mubarak Shah.
\newblock {Deep Learning Human Mind for Automated Visual Classification}.
\newblock In {\em Proceedings of the IEEE Conference on Computer Vision and Pattern Recognition (CVPR)}, pages 6809--6818, 2017.

\bibitem{sundararajan2017axiomatic}
Mukund Sundararajan, Ankur Taly, and Qiqi Yan.
\newblock Axiomatic attribution for deep networks.
\newblock In {\em International conference on machine learning}, pages 3319--3328. PMLR, 2017.

\bibitem{inceptionv3}
Christian Szegedy, Vincent Vanhoucke, Sergey Ioffe, Jonathon Shlens, and Zbigniew Wojna.
\newblock Rethinking the inception architecture for computer vision.
\newblock {\em CoRR}, abs/1512.00567, 2015.

\bibitem{takagi2022_decoding}
Yu~Takagi and Shinji Nishimoto.
\newblock High-resolution image reconstruction with latent diffusion models from human brain activity.
\newblock In {\em Proceedings of the IEEE/CVF Conference on Computer Vision and Pattern Recognition}, pages 14453--14463, 2023.

\bibitem{takagi2023improving}
Yu~Takagi and Shinji Nishimoto.
\newblock Improving visual image reconstruction from human brain activity using latent diffusion models via multiple decoded inputs, 2023.

\bibitem{tan_efficientnet_2020}
Mingxing Tan and Quoc~V. Le.
\newblock Efficientnet: Rethinking model scaling for convolutional neural networks.
\newblock In Kamalika Chaudhuri and Ruslan Salakhutdinov, editors, {\em Proceedings of the 36th International Conference on Machine Learning, {ICML} 2019, 9-15 June 2019, Long Beach, California, {USA}}, volume~97 of {\em Proceedings of Machine Learning Research}, pages 6105--6114. {PMLR}, 2019.

\bibitem{tangermann2012review}
Michael Tangermann, Klaus-Robert M{\"u}ller, Ad~Aertsen, Niels Birbaumer, Christoph Braun, Clemens Brunner, Robert Leeb, Carsten Mehring, Kai~J Miller, Gernot~R M{\"u}ller-Putz, et~al.
\newblock Review of the bci competition iv.
\newblock {\em Frontiers in neuroscience}, 6:55, 2012.

\bibitem{taylor2017cambridge}
Jason~R Taylor, Nitin Williams, Rhodri Cusack, Tibor Auer, Meredith~A Shafto, Marie Dixon, Lorraine~K Tyler, Richard~N Henson, et~al.
\newblock The cambridge centre for ageing and neuroscience (cam-can) data repository: Structural and functional mri, meg, and cognitive data from a cross-sectional adult lifespan sample.
\newblock {\em neuroimage}, 144:262--269, 2017.

\bibitem{thorpe1996speed}
Simon Thorpe, Denis Fize, and Catherine Marlot.
\newblock Speed of processing in the human visual system.
\newblock {\em nature}, 381(6582):520--522, 1996.

\bibitem{vaswani2017attention}
Ashish Vaswani, Noam Shazeer, Niki Parmar, Jakob Uszkoreit, Llion Jones, Aidan~N Gomez, {\L}ukasz Kaiser, and Illia Polosukhin.
\newblock Attention is all you need.
\newblock {\em Advances in neural information processing systems}, 30, 2017.

\bibitem{wang_image_2004}
Zhou Wang, A.C. Bovik, H.R. Sheikh, and E.P. Simoncelli.
\newblock Image quality assessment: from error visibility to structural similarity.
\newblock {\em IEEE Transactions on Image Processing}, 13(4):600--612, April 2004.
\newblock Conference Name: IEEE Transactions on Image Processing.

\bibitem{williams2020validation}
Nikolas~S Williams, Genevieve~M McArthur, Bianca de~Wit, George Ibrahim, and Nicholas~A Badcock.
\newblock A validation of emotiv epoc flex saline for eeg and erp research.
\newblock {\em PeerJ}, 8:e9713, 2020.

\end{thebibliography}
